 \documentclass[manuscript]{acmart}
 \usepackage{listings}
\usepackage[utf8]{inputenc}
    \usepackage{listings}
    \usepackage{xcolor}
    \usepackage{tikz}
\usepackage[most]{tcolorbox}
\usepackage{xcolor}

\usepackage{tabularx}
\usepackage{array}
\usepackage{ragged2e}
\usepackage{booktabs}
\usepackage{siunitx}

\usepackage{amssymb}
\usepackage{newtxmath}  % or whichever package clashes

\AtBeginDocument{%
  }

\setcopyright{acmlicensed}
\copyrightyear{2018}
\acmYear{2018}
\acmDOI{XXXXXXX.XXXXXXX}
\acmConference[Conference acronym 'XX]{Make sure to enter the correct
  conference title from your rights confirmation email}{June 03--05,
  2018}{Woodstock, NY}
\acmISBN{978-1-4503-XXXX-X/2018/06}

\begin{document}

%%
%% The "title" command has an optional parameter,
%% allowing the author to define a "short title" to be used in page headers.
\title[Pedagogical Chatbots for Instructors]{Exploring Conversational Design Choices in LLMs for Pedagogical Purposes: Socratic and Narrative Approaches for Improving Instructor's Teaching Practice}

\author{Si Chen}
\affiliation{%
  \institution{University of Notre Dame}
  \city{Notre Dame}
  \state{Indiana}
  \country{United States}}
\email{schen34@nd.edu} % update if different

\author{Isabel R. Molnar}
\affiliation{%
  \institution{University of Notre Dame}
  \city{South Bend}
  \state{Indiana}
  \country{United States}}
\email{imolnar@nd.edu} % add if available

\author{Peiyu Li}
\affiliation{%
  \institution{University of Notre Dame}
  \city{Notre Dame}
  \state{Indiana}
  \country{United States}}
\email{pli9@nd.edu}

\author{Adam Acunin}
\affiliation{%
  \institution{University of Notre Dame}
  \city{Notre Dame}
  \state{Indiana}
  \country{United States}}
\email{aacunin@nd.edu} % add if available

\author{Ting Hua}
\affiliation{%
  \institution{University of Notre Dame}
  \city{Notre Dame}
  \state{Indiana}
  \country{United States}}
\email{thua@nd.edu} % add if available

\author{Alex Ambrose}
\affiliation{%
  \institution{ University of Notre Dame}
  \city{Notre Dame}
  \state{Indiana}
  \country{United States}}
\email{aambrose@nd.edu} % add if available

\author{Nitesh V. Chawla}
\affiliation{%
  \institution{University of Notre Dame}
  \city{South Bend}
  \state{Indiana}
  \country{United States}}
\email{nchawla@nd.edu} % add if available

\author{Ronald Metoyer}
\affiliation{%
  \institution{University of Notre Dame}
  \city{South Bend}
  \state{Indiana}
  \country{United States}}
\email{rmetoyer@nd.edu} % add if available

%%
%% The "author" command and its associated commands are used to define
%% the authors and their affiliations.
%% Of note is the shared affiliation of the first two authors, and the
%% "authornote" and "authornotemark" commands
%% used to denote shared contribution to the research.

%%
%% By default, the full list of authors will be used in the page
%% headers. Often, this list is too long, and will overlap
%% other information printed in the page headers. This command allows
%% the author to define a more concise list
%% of authors' names for this purpose.
\renewcommand{\shortauthors}{Trovato et al.}

%%
%% The abstract is a short summary of the work to be presented in the
%% article.
\begin{abstract}
Large language models (LLMs) typically generate direct answers, yet they are increasingly used as learning tools. Studying instructors’ usage is critical, given their role in teaching and guiding AI adoption in education. We designed and evaluated \textit{TeaPT}, an LLM for pedagogical purposes that supports instructors’ professional development through two conversational approaches: a \textit{Socratic} approach that uses guided questioning to foster reflection, and a \textit{Narrative} approach that offers elaborated suggestions to extend externalized cognition. In a mixed-method study with 41 higher-education instructors, the \textit{Socratic} version elicited greater engagement, while the \textit{Narrative} version was preferred for actionable guidance. Subgroup analyses further revealed that less-experienced, AI-optimistic instructors favored the \textit{Socratic} version, whereas more-experienced, AI-cautious instructors preferred the \textit{Narrative} version. We contribute design implications for LLMs for pedagogical purposes, showing how adaptive conversational approaches can support instructors with varied profiles while highlighting how AI attitudes and experience shape interaction and learning.
\end{abstract}

%%
%% The code below is generated by the tool at http://dl.acm.org/ccs.cfm.
%% Please copy and paste the code instead of the example below.
%%
\begin{CCSXML}
<ccs2012>
   <concept>
       <concept_id>10003120.10003121</concept_id>
       <concept_desc>Human-centered computing~Human computer interaction (HCI)</concept_desc>
       <concept_significance>500</concept_significance>
       </concept>
   <concept>
       <concept_id>10010405.10010489.10010490</concept_id>
       <concept_desc>Applied computing~Computer-assisted instruction</concept_desc>
       <concept_significance>500</concept_significance>
       </concept>
   <concept>
       <concept_id>10003456.10003457.10003527</concept_id>
       <concept_desc>Social and professional topics~Computing education</concept_desc>
       <concept_significance>500</concept_significance>
       </concept>
 </ccs2012>
\end{CCSXML}

\ccsdesc[500]{Human-centered computing~Human computer interaction (HCI)}
\ccsdesc[500]{Applied computing~Computer-assisted instruction}
\ccsdesc[500]{Social and professional topics~Computing education}
%%
%% Keywords. The author(s) should pick words that accurately describe
%% the work being presented. Separate the keywords with commas.
\keywords{Instructors, Socratic, AI Attitude}
%% A "teaser" image appears between the author and affiliation
%% information and the body of the document, and typically spans the
%% page.
% \begin{teaserfigure}
%   \includegraphics[width=\textwidth]{sampleteaser}
%   \caption{Seattle Mariners at Spring Training, 2010.}
%   \Description{Enjoying the baseball game from the third-base
%   seats. Ichiro Suzuki preparing to bat.}
%   \label{fig:teaser}
% \end{teaserfigure}

\received{20 February 2007}
\received[revised]{12 March 2009}
\received[accepted]{5 June 2009}

%%
%% This command processes the author and affiliation and title
%% information and builds the first part of the formatted document.
\maketitle

\section{Introduction}

%\textbf{Promise.} 
Artificial intelligence (AI) tools, particularly large language models (LLMs), are increasingly positioned as companions for learning and professional development. Their promise lies in delivering personalized support: helping learners of all kinds, from students to working professionals, reflect, practice, and grow at scale. If designed well, such systems could extend opportunities for lifelong learning by not only providing efficient access to information but also supporting deeper processes of reasoning, reflection, and skill development \cite{luckin2016intelligence}. 

%\textbf{Obstacles.} 
Despite this promise, two major obstacles remain.  
First, most LLMs, by design, generate direct answers rather than facilitate teaching and learning. This makes them efficient but not inherently pedagogical: they excel at delivering solutions but rarely foster understanding \cite{koedinger2012kli,collins2018rethinking}. Recent product efforts, such as ChatGPT’s \textit{Study Mode} \cite{openai2025studymode} introduced in July 2025, signal movement toward learning-focused interactions through socratic-style questioning, yet it remains unclear whether such approaches can meaningfully support  learning.
% [adding why we should design for teachers]-- how they are playing a key role in how AI should be used in teaching and classrooms 

Second, instructor-facing generative AI tools are proliferating, such as syllabus generators, assignment creators, and grading assistants, but empirical studies of how instructors actually engage with these tools remain limited. Prior work shows that AI literacy and attitudes are becoming critical barriers: some instructors experiment actively, while others remain skeptical \cite{chen2025bridging}. Because instructors are responsible not only for adopting AI themselves but also for setting norms and serving as role models in how AI should be used, this heterogeneity creates uneven capacity to address students’ AI practices \cite{luckin2016intelligence,kamali2024ai}. 
Instructors’ choices and behaviors ripple outward, influencing how students interpret what counts as responsible or effective use. This underscores the need for empirical evidence on how design choices in AI systems intersect with instructors’ professional backgrounds and responsibilities.

%\textbf{Our approach.} 
We address these challenges by designing and evaluating \textit{TeaPT}, an LLM-based prototype with two conversational approaches for teaching pedagogical strategies. The \textit{Socratic} approach emphasizes reflective questioning, aligning with theories of dialogic and interactive learning that highlight the value of reasoning through guided dialogue \cite{chi2014icap,wegerif2007dialogic,mercer2000words}. The \textit{Narrative} approach emphasizes elaboration and breadth, aligning with theories of externalized cognition, where knowledge is extended through rich external suggestions and representations \cite{hutchins1995cognition,clark1998extended,scardamalia2006}. %These approaches represent distinct but complementary cognitive processes: socratic questioning fosters internalized reasoning, while elaborative narration broadens externalized cognition. 
Together, they offer multiple processes for professional learners to engage with LLMs. In a mixed-method study with 41 higher-education instructors, we examined how these two approaches shaped engagement, survey ratings, and qualitative perceptions, including variation by AI attitude and teaching experience.

\textbf{Summary of contributions.}  
We empirically show how conversational approach (\textit{socratic} vs. \textit{narrative}) shapes engagement and reflective learning opportunities for professional learners, specifically higher education instructors.  
We demonstrate how instructors’ backgrounds mediate these effects. For example, the \textit{socratic} version was especially useful for junior instructors and those more optimistic about AI.  
We contribute design implications for LLMs for pedagogical purposes.%: \textit{socratic} questions foster dialogic reflection but risk inefficiency, while \textit{narrative} lists extend external cognition but risk overload, motivating adaptive, profile-sensitive designs.  

\section{Related Works} 

\subsection{LLMs for Learning and Tutoring}

The use of AI as a teaching and learning technology has a long history in education, most notably through intelligent tutoring systems (ITS) designed to foster reasoning, practice, and reflection rather than simply deliver answers \cite{koedinger2012kli,collins1991,chi2014icap,luckin2016}. These systems emphasized scaffolding through guided practice and feedback, with frameworks such as cognitive apprenticeship \cite{collins1991} and ICAP \cite{chi2014icap} showing that active engagement and reflection are critical to durable learning. This work established AI not only as a source of information but also as a partner that shapes how learners reason and practice.  

The LLM has more recently been studied as a teaching and learning technology with mixed results. EducationQ \cite{shi2025educationq} evaluates instructional effectiveness across models, showing that teaching quality does not scale linearly with size; PACE \cite{liu2025one} simulates learner profiles and adapts Socratic feedback to personalize mathematics tutoring; and Agent4Edu \cite{gao2025agent4edu} uses profile and memory modules to simulate learner responses for large-scale evaluation. These studies highlight the potential of LLMs for interaction and personalization, yet most LLMs remain optimized to provide direct answers rather than to scaffold reasoning or reflection. Their actual effects on learning outcomes, and the design choices needed to make them pedagogically effective, remain open questions.  

At the same time, generative AI applications such as ChatGPT are already being adopted as teaching and learning tools across contexts. Students use them for studying and homework, often as a supplement to traditional resources \cite{tang2025understanding}. Instructors adopt them for lesson preparation and teaching support, though surveys report ambivalence and uneven patterns of uptake \cite{pew2024,mamo2024,cabellos2024}. Beyond formal education, adults in the workforce are using generative AI for professional learning, such as developing new skills, practicing workplace communication, and pursuing upskilling and reskilling opportunities \cite{pew2025workforce,luckin2022ai}. These trends show that generative AI is making learning accessible to a wide range of learners, while also raising new challenges and opportunities for the design of systems that act not only as problem solvers but as effective pedagogical partners. This motivates our work on \textit{TeaPT}, which examines how conversational design choices can make LLMs more effective for learning.  

\subsection{Instructor-Facing AI Tools and Literacy}

Instructors are important professional learners: they use AI in their own teaching and also guide how students engage with it. Recent years have seen a rise in instructor-facing generative AI tools such as syllabus generators, grading assistants, and assignment creators. However, studies on how instructors actually use these tools are still limited. Early surveys show mixed views: K--12 teachers report ambivalence about classroom AI \cite{pew2024}, while higher-education faculty show mostly neutral-to-positive attitudes and gradual adoption \cite{mamo2024,cabellos2024}. These patterns align with broader models of technology adoption such as UTAUT \cite{venkatesh2003}, which emphasize factors like usefulness, effort, social influence, and available support. Yet the lack of detailed evidence leaves open how design choices can better meet instructors’ professional learning needs.  

Research also shows that instructors’ AI skills, attitudes, and responsibilities strongly shape adoption. Some experiment actively, while others remain skeptical, a divide consistent with broader findings on algorithm aversion and appreciation \cite{dietvorst2015,dietvorst2015_pub,logg2019}. Work in human–automation emphasizes trust as a central factor in deciding when to rely on AI, with reviews detailing what influences trust and how it can be maintained \cite{lee2004,hoff2015}. HCI studies add that complementary performance depends on users’ mental models and calibrated reliance, and that design patterns such as cognitive forcing functions can reduce overreliance \cite{bansal2021,buccinca2021trust}. For instructors, this means that attitudes and prior experience will directly affect how they respond to different AI interaction styles.  

Design guidelines in HCI emphasizes that human–AI systems should anticipate, support, and adapt to diverse users rather than assume one-size-fits-all interactions \cite{amershi2019}. In parallel, AI literacy work argues for scaffolding competencies so users can engage critically and productively with AI \cite{long2020}. Taken together, these studies show that instructors differ widely, some being early adopters and others skeptical, and highlight the importance of supporting them both as professional learners and as leaders in shaping students’ practices. This motivates our work on \textit{TeaPT}, which explores how different conversational designs can adapt to instructor profiles and better support professional learning.

\subsection{Dialogic and External Cognition Processes}

Our research is informed by two theoretical perspectives that have been widely applied in education and HCI research: dialogic learning and external cognition. We focus on these perspectives because they correspond to two prevalent forms of interaction in conversational approaches (short reflective questions and extended lists) and have been shown to support distinct learning processes. While other theories address related aspects of learning, such as motivation or self-regulation, dialogic and external cognition frameworks are particularly relevant here because they directly capture the tradeoff between reflection and efficiency that emerges in Human-LLM interaction.

\textbf{Dialogic learning.} Socratic questioning builds on traditions of dialogic teaching, where interaction and reasoning are prioritized over the direct transmission of information. The ICAP framework shows that interactive and constructive activities, such as explaining concepts to others, generating examples, or answering guiding questions, lead to more durable learning than passive reception \cite{chi2014icap}. Dialogic teaching similarly highlights how questions that encourage learners to articulate and reason, for instance through debate, collaborative problem-solving, or reflective questioning, can deepen understanding and support transfer \cite{wegerif2007dialogic}. HCI and CSCW studies have drawn on these perspectives to design systems such as collaborative learning platforms and tutoring interfaces that engage users in active reflection rather than immediate solution-finding, underscoring the broader relevance of dialogic principles in technology-mediated learning.  

\textbf{External cognition.} A complementary perspective emphasizes how cognition extends through external representations and artifacts. Clark and Chalmers’ theory of the extended mind \cite{clark1998extended} and Hutchins’ work on distributed cognition \cite{hutchins1995cognition} argue that reasoning and memory are often supported by external tools, such as notes, diagrams, or shared displays, that make knowledge more accessible and actionable. In education, knowledge-building theory highlights the importance of elaborated external contributions. For example, learners building on each other’s written ideas in online forums for collective and individual growth \cite{scardamalia2006}. These ideas have informed HCI research on systems such as visualization tools, collaborative document editors, and shared dashboards that scaffold decision-making and support collaborative knowledge construction, illustrating how externalized representations can broaden access to strategies while also raising concerns about information overload.

\section{Formative Study} \label{sec:formative}

\subsection{Process}
%We conducted a two-stage formative study to explore how pedagogy experts perceive AI-driven instructional support. 
We conducted a two-stage formative study to understand instructor-support strategies and how pedagogy expert perceive AI-driven instructional support. pedagogy expert
\textbf{Activity 1: Expert Interviews} We first interviewed five pedagogy experts to understand current strategies used to support instructors and to surface opportunities and risks associated with AI. A pedagogy expert has deep expertise in guiding instructors, especially around AI and broader technology adoption. Each interview lasted approximately 60 minutes. \textbf{Activity 2: Design Session and Evaluation} We then conducted a design session with ten pedagogy experts, who reviewed a storyboard of chatbot features tailored to instructors with varying levels of AI literacy and differing attitudes toward AI. Each session lasted about 60 minutes. Experts also evaluated LLM-generated suggestions for 20 common instructional challenges curated from actual instructors. The goal of this rating task was to assess how well ChatGPT could support teachers and to identify areas for improvement. Suggestions were rated on a 1–10 scale (1 = worst, 10 = best) across four dimensions, informed by participants' design session feedbacks: Relevance, Actionability, Alignment, and Reliability. Results showed mean scores of 7.05, 6.22, 6.76, and 6.76, respectively, indicating both promising performance and opportunities for further refinement. Interview and design session transcripts were thematically analyzed to identify recurring concerns and suggestions. Two researchers independently coded the transcripts and refined themes through discussion until consensus was reached, following Braun and Clarke’s guidelines for thematic analysis \cite{braun2006using}.

\subsection{Design Insights}
\textbf{D1: Accommodating a Wide Range of Instructors’ AI Attitudes and Readiness} Pedagogy experts noted a wide spectrum of instructor readiness for AI adoption, ranging from early adopters to those highly cautious of technological change. To accommodate this spectrum, AI systems should incorporate clear mechanisms for transparency and control, such as options to track, manage, and delete data. Two pathways emerged as particularly salient: onboarding processes that establish initial preferences, and ongoing conversation logs that can be curated or removed through a dashboard. Instructors also expressed greater willingness to engage with AI when suggested changes to teaching practices were modest and incremental. Because AI literacy and attitudes are closely intertwined, and because instructors often seek support in areas beyond their expertise, sweeping shifts in pedagogy frequently generated resistance. Design efforts should therefore emphasize low-barrier entry points and gradual integration into existing instructional practices.

\textbf{D2: Emphasizing Scaffolding Rather than Direct Solutions} Participants highlighted a tension between AI’s ability to provide quick answers and its usefulness for professional learning. While tools like ChatGPT were valued for fact checking, pedagogy experts noted that direct solutions offered little opportunity for reflection or growth. They preferred questions that encouraged deeper engagement with their own reasoning.  To address this, AI systems should emphasize scaffolding strategies that foster multi-turn reflection. Contextualized questions, rather than final answers, allow instructors to articulate their thinking, test alternatives, and refine their practices.  We explored this contrast in our prototype by comparing user experience when interacting with the standard ChatGPT API versus a fine-tuned model that was tuned to promote reflection and metacognitive engagement.

\textbf{D3: Fostering Calibrated Trust in LLMs’ Pedagogical Use} Pedagogy experts emphasized that while LLMs are powerful, their pedagogical grounding and domain expertise remain uncertain. The aim is not to simply increase or decrease trust, but to align it with the system’s actual capabilities and limitations. In higher education, even courses with the same title often differ in emphasis and design, making pedagogy experts cautious about AI-generated domain knowledge. To support calibrated trust, systems should avoid presenting themselves as authoritative experts. Rather than producing course-specific content upfront, they should prompt instructors to draw on their own expertise while grounding suggestions in trusted, verifiable sources. In this way, AI functions as a proxy that connects instructors to reliable resources and opportunities for human collaboration, rather than as a substitute for their or all human expertise.

\begin{figure}
    \centering
    \includegraphics[width=10cm]{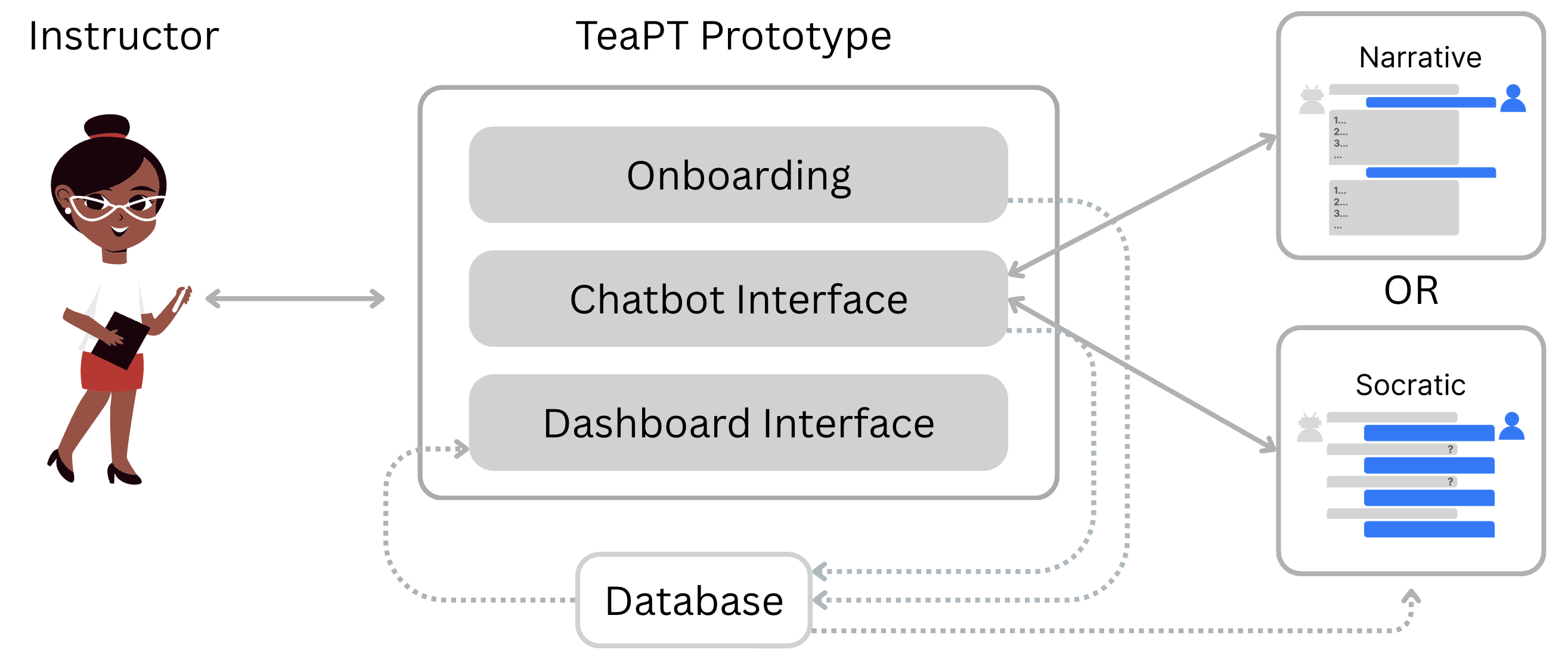}
    \caption{The TeaPT web interface is composed of three main components: the onboarding, chatbot interface, and dashboard. These components all interact with the user, conversation agent, and database.}
    \label{fig:system_structure}
\end{figure}

\section{TeaPT Prototype Design and Development}

\subsection{Prototype Overview}

We introduce TeaPT (Teaching Excellence and Pedagogical Transformation), an LLM-based chatbot designed to enhance teaching through a conversational interface. The TeaPT prototype was designed based on insights collected from the formative study. The prototype’s architecture includes three core components, as illustrated in Fig. \ref{fig:system_structure}: a one-time onboarding process, a chatbot interface, and a user dashboard. TeaPT utilizes  two approaches: a \textit{narrative} version, which provides elaborated strategies and resources, and a \textit{socratic} version, which offers concise, reflective questions to scaffold improvement rather than directly supplying solutions, aligning with the scaffolding needs identified in (D2).

The initial user interaction begins with the onboarding process, completed upon the user’s first login. This process collects foundational user data through four initial questions, with the option for users to skip questions they are uncomfortable answering. All data collected at this stage is stored in the user’s profile in the database. After onboarding, users are directed to the chatbot interface, where they are first guided through a walkthrough. The primary mode of interaction is a text box (F1), with all conversational history visible in the sidebar. Users can also associate a challenge with a specific course by selecting it via the plus button (F2). 

As shown in Fig. \ref{fig:system_structure}, the data collected during the onboarding is utilized by the conversational agent to provide a personalized experience. To address the challenge of initial user hesitancy (D1) and foster trust in the LLM, the agent is designed to progressively ask the user about their instructional background. This gradual data collection strategy minimizes the initial data burden on the user. The subsequently collected information is then stored in the database.

The final component is the user dashboard, detailed in Appendix \ref{app:dashboard}, which serves as a centralized hub for managing the TeaPT experience. From the dashboard, users can access their teaching resources, review demographic information that was extracted from their conversation with the prototype, and identify course-specific challenges. They can also schedule live consultations with human pedagogy experts, in order to calibrate trust and connect with other verifiable resources (D3). After completing each conversation with the conversational agent, the TeaPT prototype leverages ChatGPT to summarize the user’s primary challenge from the interaction and create a tailored resource to help them address it, these data are then saved to the dashboard for easier access.

\begin{figure}[htp]
    \centering
    \includegraphics[width=11cm]{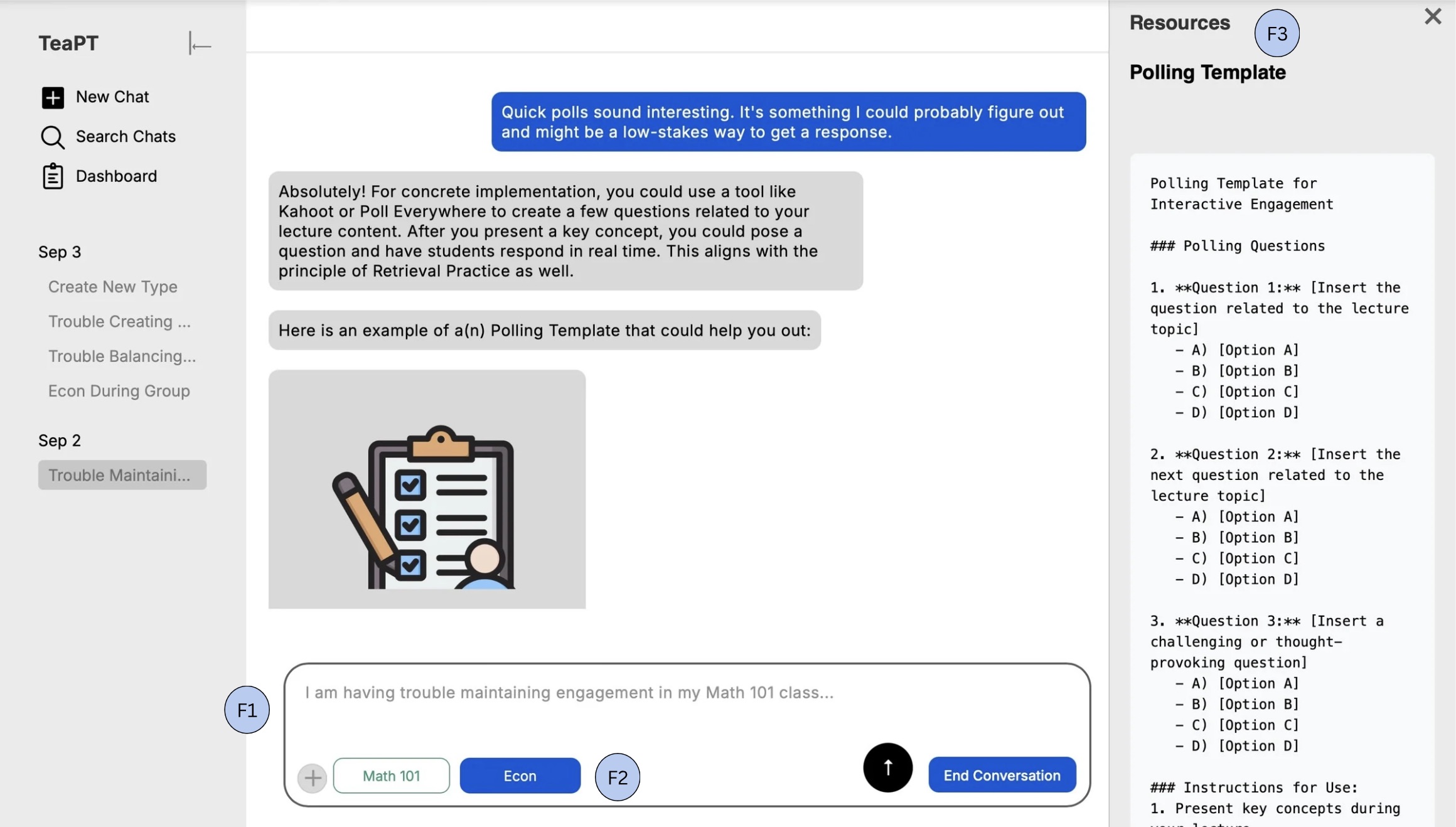}
    \caption{The TeaPT chatbot interface allows users to interact with the conversational agent. After conversation completion, the prototype produces a resource to help instructors implement feedback in their classroom.}
    \label{fig:system_overview}
\end{figure}

\subsection{Chatbot Interface - Two Conversational Approaches}

The \textit{TeaPT} prototype incorporates two different conversational approaches designed to support professional learning: a \textit{socratic} version and a \textit{narrative} version, as shown in Fig. \ref{fig:system_overview}. The prototype was deliberately scoped to focus on core instructional challenges, such as classroom management, student engagement, and assessment design, rather than broader institutional concerns (e.g., policy, academic integrity, or the governance of AI in education).

The \textit{socratic} version is driven by a fine-tuned Llama model \cite{touvron2023llama} and is designed to impart knowledge through a series of questions. This approach prioritizes depth of knowledge and encourages user reflection. The \textit{socratic} Version is characterized by shorter, more frequent conversational turns, fostering a dynamic exchange and reinforcing concepts through iterative dialogue. 

In contrast, the \textit{narrative} version is powered by calls to ChatGPT and leverages stories and examples to deliver information. This version prioritizes the breath and context of a topic. The \textit{narrative} version generates significantly longer and more detailed responses, which can be more challenging for users to fully process and comprehend.

\subsubsection{Socratic Version} We fine-tuned the \texttt{meta-llama/Llama-2-13b-chat-hf} model using a structured \textit{socratic} dataset consisting of 406,183 chat examples (user--assistant pairs), along with a held-out validation set of 4,156 examples. Each example began with a standardized system prompt defining the assistant's role as an expert teaching consultant (see Appendix~\ref{app:socratic-prompt}). 

To support structured, step-by-step pedagogical reasoning, we introduced two special tokens, \texttt{<step>} and \texttt{</step>}, into the training data. These tokens enclosed a short descriptor that explicitly marked the assistant’s current instructional phase—for example, \texttt{<step>Current Step: Step 1</step>}. Before generating a response, the model was expected to recognize this step label and condition its output based on the corresponding guidance provided in the system prompt (see Appendix~\ref{app:socratic-prompt}). For instance, if the label was \texttt{<step>Current Step: Step 1</step>}, the model focused on identifying the user’s instructional challenge; if the label indicated Step 2, it aimed to explore reasons underlying that challenge. This explicit step-labeling mechanism enabled the model to scaffold its responses across dialogue turns in a structured and pedagogically coherent manner. During training, the model learned to associate each step with distinct instructional behaviors, internalizing both the structure and intent of Socratic teaching interactions (see Appendix~\ref{app:inst-template}). As a result, the model demonstrated a clearer understanding of instructional flow and was able to guide users through complex teaching scenarios using expert-like scaffolding. Additional details about the training configuration—such as hardware setup, optimizer parameters, and learning schedule—are provided in Appendix~\ref{app:training-details}.

Our scaffold adapts a three-step problem-solving framework—identify the problem, explore reasons, and develop strategies—into an expert–instructor dialogue \cite{cmuSolveProblem}. We added an initial step to greet users and gather background before moving to diagnosis and tailored strategies. The dialogues are synthetic and grounded in Small Teaching \cite{lang2016small} and Small Teaching Online \cite{darby2019small}, two of the most widely adopted faculty development resources in higher education in the US. Small Teaching has sold over 100,000 copies and is regularly used in faculty workshops and teaching centers, while Small Teaching Online has been adopted at over 200 institutions, particularly following the COVID-19 pandemic. To ensure validity, the generated synthetic data underwent three rounds of expert review (50 conversation per round), involving the books’ authors as well as additional pedagogy experts, confirming alignment with the pedagogical intent of the original texts. The detailed procedures for synthetic data generation and validation are beyond the scope of this paper and will be presented in a separate publication.

% \begin{figure}[htp]
%     \centering
%     \includegraphics[width=12cm]{figure/Onboarding.jpeg}
%     \caption{The TeaPT web interface is composed of three main components: the onboarding, chatbot interface, and dashboard. These components all interact with the user, conversation agent, and database in order to shape the TeaPT UI/UX.}
%     \label{fig:system_overview}
% \end{figure}

\subsubsection{Narrative Version} 
The \textit{narrative} version was implemented through API calls to ChatGPT. The prompt, detailed in Appendix \ref{app:chat-prompt}, incorporated the user’s instructional background and instructed the model to respond as an experienced teaching expert, for example, “You are an experienced teaching expert guiding instructors to solve teaching challenges based on concepts and examples.” This framing consistently produced extended, information-rich responses, often structured as lists or step-by-step strategies.  

%As part of our design process, we first tested the same prompt with both non–fine-tuned LLaMA and ChatGPT. In pilot testing, however, LLaMA’s default conversational approaches did not generate the structured, list-like formatting we required and lacked the level of teaching-specific knowledge observed in the \textit{socratic} version. This outcome was expected, as fine-tuning typically alters the knowledge space itself, making direct comparison unfair. ChatGPT, by contrast, demonstrated adequate teaching knowledge out of the box and, with the same prompt framing, reliably produced the structured and elaborated outputs we desired.   

\subsubsection{Teaching Resource Generation (Same in the two versions)} The TeaPT prototype also includes a resource generator feature, which is activated towards the end of a conversation. This feature provides a tailored resource, such as a rubric or a survey template, to help the instructor implement the strategies discussed. These resources are created through an API call to ChatGPT, utilizing the prompt detailed in Appendix \ref{app:resource-prompt}. When a resource is generated, users can click to open and view it within the resource window (F3). This allows users to easily review the content before saving it to their personalized dashboard, where all collected resources are accessible (F5).

\subsection{Pilot Testing}

After the creation of the TeaPT prototype, the research team held a pilot study with six experienced pedagogy experts in July 2025, different from those that participated in the formative study. Each expert participated in a two to three-hour session, during which they evaluated the prototype’s usability and the quality of its responses for both the \textit{socratic} version and \textit{narrative} version. The experts were compensated 40 USD per hour for their time. Based on the feedback from the pilot study, the research team was able to make minor design modifications, refine the rating scale used in the later user study (see Appendix~\ref{survey}), and adjust the API prompts to better align with the users’ needs.

\section{User Study and Results} 

\subsection{Study Design} The research team conducted a mixed-method Zoom study in August 2025. Each participant interacted with two versions of our tool for 10 minutes each: one with the core QA in the \textit{narrative} version and the other in the \textit{socratic} version, with only the core QA component differing, seeking suggestions on the same initial questions under both versions. The order of exposure was counterbalanced across participants. After interacting with each version, participants rated the version they had just used, followed by a 15-minute post-study interview. Two co-authors alternated in leading the study. We address the following research questions: \textit{\textbf{RQ1}: How do instructors perceive and engage with Socratic versus Narrative pedagogical chatbot?} \textit{\textbf{RQ2}: How do instructor characteristics (e.g., AI attitudes, teaching experience) influence preferences and perceived learning across the two versions?}

\subsubsection{Participants Recruitment and Demographics} 
We recruited 41 higher education instructors through social media and public posts. Recruitment was open and not selective, drawing individuals from a wide range of U.S. institutions, including research universities, community colleges, and liberal arts colleges. Demographic information was self-reported during sign-up and screening and further confirmed during the Zoom study sessions. The sample included 22 current faculty and 19 teaching assistants or staff. Participants varied in years of teaching experience: 1--2 years (n=13, 31.7\%), 10+ years (n=11, 26.8\%), 0--1 year (n=8, 19.5\%), 3--5 years (n=5, 12.2\%), and 5--10 years (n=4, 9.8\%). Attitudes toward AI were nearly evenly divided, with 21 participants identifying as \textit{Open and Optimistic} and 20 as \textit{Curious but Cautious}. Although the sign-up form included an additional option (\textit{Not for me}), no participant selected it, indicating that our sample only represents individuals who are at least somewhat open to AI. With informed consent obtained at the start, each session lasted approximately 45 minutes in total, and each participant received 35 USD as compensation. There is no evidence of association between AI attitude and Role using a Chi-square test. Similarly, there is no significant association between YOE (0-2/3+) and AI attitude (\textit{Curious but Cautious} vs \textit{Open and Optimistic}).

\subsubsection{Within-Subject Study Procedure Details and Analysis} Each participant interacted with both versions of the tool sequentially within a single session and completed two surveys right after they interacted with each version. The onboarding process was repeated for each version, with only the core question–answering component differing. This design allowed participants to more clearly understand and evaluate the differences between the two versions. Participants were told that there were two versions but were not informed which was which or how they differed.  Participants were encouraged to think-aloud during the interactions with each version. Participants were asked to identify a current teaching challenge they wished to improve upon and interact with the chatbot to seek possible solutions. Interaction with each tool version was set to 10 minutes, based on pilot testing which indicated that this duration was sufficient to meaningfully discuss at least one teaching challenge. Participants could end the session earlier if they wished, were not required to use the entire time, and were free to raise multiple challenges as long as they completed discussion of at least one.

\textbf{Conversation Log} To answer RQ1, we conducted within-subject paired tests across three engagement metrics: \textit{User Message Count}, \textit{User Word Count}, and \textit{Chatbot Response Word Count}, comparing the \textit{socratic} version and the \textit{narrative} version for each participant with valid data (N = 39 with full data save successfully). Normality of the paired differences was assessed using the Shapiro–Wilk test. If differences were approximately normal, we ran a paired t-test. If not, we ran a Wilcoxon signed-rank test. Additionally to answer RQ2, we compared paired differences in the outcome measures using within-subject tests (paired t-tests when normality held, Wilcoxon signed-rank otherwise). To examine whether these differences varied across participant subgroups to answer RQ2, we then compared the difference scores between groups using either Welch’s two-sample t-tests or Wilcoxon rank-sum tests, depending on distributional assumptions. Multiple comparisons within each subgrouping were corrected using Holm’s adjustment. To assess the sensitivity of our design, we conducted a power analysis using the paired $t$-test approximation, which indicated that with 39 participants data logged successfully, the study had more than 80\% power to detect medium effects (Cohen’s $d \geq 0.50$). 

\textbf{Surveys for Each Version}
The survey instrument filled in after each version contained the same set of evaluation questions developed with pedagogy experts. We chose not to adopt existing usability scales, as experts did not agree on their validity for this instructional context. Instead, participants rated both versions on five pedagogically grounded dimensions (1–3 scale for worst to best; (see Appendix~\ref{survey})): \textit{Clarity of Expression}, \textit{Supportive \& Appropriate Tone}, \textit{Appropriateness of Validation}, \textit{Reflective Prompting}, and \textit{Actionable Guidance}.  
To answer RQ1, we compared paired ratings between the two versions on each of the five dimensions using Wilcoxon signed-rank tests, treating the three-point scales as ordinal. For each dimension, we report the test statistic , $p$-value, 95\% confidence interval, and the median paired difference (positive values indicating higher ratings for our model). %To assess the sensitivity of our design, we conducted a power analysis using the paired $t$-test approximation, which indicated that with 41 participants the study had 80\% power to detect medium effects (Cohen’s $d \geq 0.45$).

To examine whether instructors’ backgrounds shaped their perceptions of the two versions (RQ2), we analyzed variation across three background factors: \textit{AI attitude}, \textit{teaching experience}, and \textit{faculty status}. AI attitude was measured on the sign-up form, where participants identified themselves as either \textit{Open and Optimistic} or \textit{Curious but Cautious}. Teaching experience was dichotomized as junior \textit{0--2 years} versus seasoned \textit{3+ years}. Faculty status distinguished between \textit{current faculty} and \textit{teaching assistants/staff}. For each evaluation dimension, we computed a within-participant difference score (\textit{socratic} $-$ \textit{narrative}, range $-2$ to $+2$). A positive score indicated a preference for \textit{socratic}, a negative score indicated a preference for \textit{narrative}, and a zero indicated no preference. To test whether participants in each subgroup demonstrated systematic preferences, we applied two complementary nonparametric tests: the \textit{Wilcoxon signed-rank test}, which evaluates whether the median difference score differs from zero, and the \textit{binomial sign test}, which compares the number of participants preferring each version while discarding ties. In addition, to examine whether background groups differed from one another, we applied between-group tests: \textit{Fisher’s exact test} on non-tied preference counts (socratic vs.\ narrative) and the \textit{Wilcoxon rank-sum test} on the continuous difference scores. To assess adequacy of our sample, we conducted an a priori--style power analysis using paired $t$-test approximations (equivalent to our within-participant signed-rank and sign tests). With $n = 41$ paired observations, we had more than $80\%$ power to detect medium effects (Cohen’s $d \geq 0.50$) at $\alpha = .05$. Thus, our within-subject preference analyses were sufficiently powered to identify meaningful differences between versions. Multiple comparisons within each subgrouping were corrected using Holm’s adjustment.

\subsubsection{Post Study Interview}
Towards the end of the study, participants completed a semi-structured interview designed to elicit reflections on their experience with the two chatbot versions and overall experience. Questions probed perceived differences between the \textit{narrative} and \textit{socratic} versions, reasons for preferences, and how participants’ backgrounds (e.g., teaching experience, AI attitudes) may have shaped their perceptions. To examine potential instructional impact, we also asked an application-oriented question: \textit{“What did you learn today by talking to the two chatbots that you would change or use in your next teaching session, and how did you learn that?”}. We treated these responses as a proxy for learning outcomes, linking perceptions of the chatbots to potential instructional practices. Interview transcripts were analyzed thematically using an inductive coding approach, with two co-authors independently coding and iteratively reconciling differences to ensure agreement. For the application-oriented responses, the same coders further classified whether participants attributed their learning to the \textit{narrative} version, the \textit{socratic} version, both, or if attribution was unclear, allowing us to assess how each version contributed to perceived instructional takeaways.

\subsection{Findings- RQ1- Differences Between Socratic and Narrative}

\begin{figure}[t]
  \centering
  \includegraphics[width=0.9\textwidth]{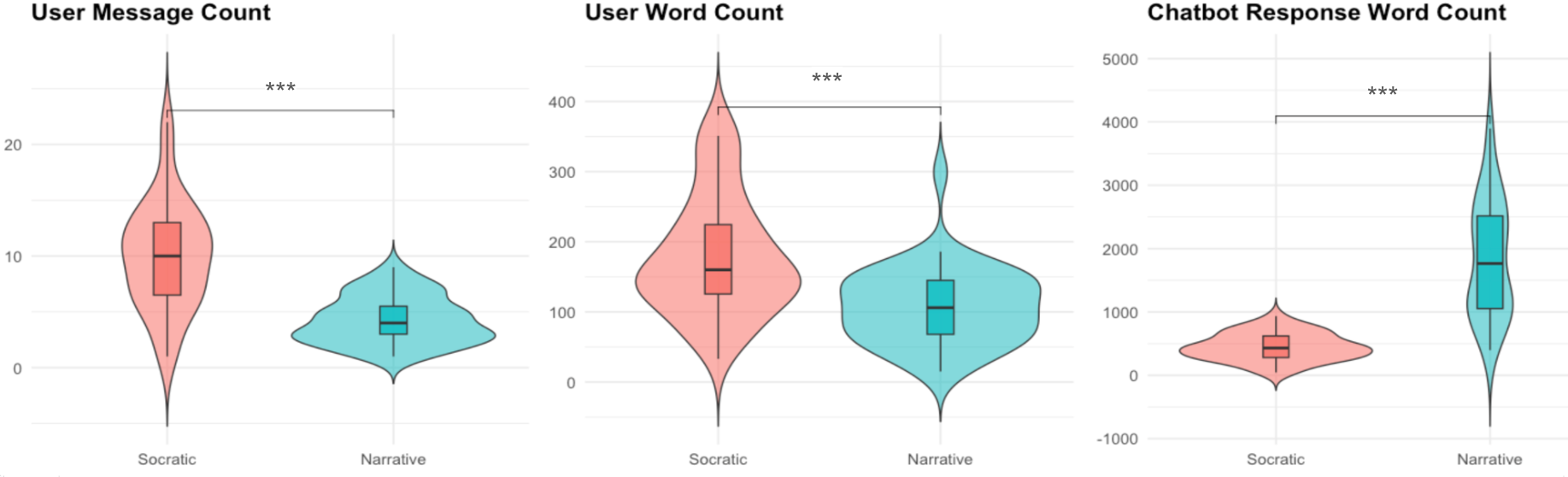}
  \caption{Distributions of user and chatbot engagement metrics across the \textit{Socratic} and \textit{Narrative} versions. Participants contributed more messages and words in the \textit{Socratic} version, consistent with its dialogic design, while the \textit{Narrative} chatbot produced substantially longer responses, consistent with its externalization focus. Significance markers indicate paired test results. (*** indicates p < 0.001)}
  \Description{}
  \label{violin}
\end{figure}

\subsubsection{Conversation Log} Comparison shows \textit{Socratic} elicited greater user engagement, reflected in higher message and word counts; while \textit{Narrative} generated much longer assistant responses as shown in Fig. 
\ref{violin}. Side-by-side examples can be seen in Appendix~ \ref{example}. 

\textit{User Message Count}
Participants engaged in significantly more conversations turns in the \textit{socratic} version 
($Mdn = 10.0$, $M = 10.03$, $SD = 4.81$) 
compared to the \textit{narrative} version 
($Mdn = 4.0$, $M = 4.23$, $SD = 2.15$). 
A Wicoxon signed-rank test indicated this difference was highly significant 
($p < .001$, Holm $p < .001$), with a median increase of $+6$ messages 
in the \textit{socratic} version.

\textit{User Word Count}
Participants also wrote substantially more words in the \textit{socratic} version 
($Mdn = 160$, $M = 178.74$, $SD = 85.33$) 
than in the \textit{narrative} version 
($Mdn = 106$, $M = 107.21$, $SD = 54.99$). 
Differences were normally distributed (Shapiro–Wilk $p = .32$), so a paired-samples 
$t$-test was used. The analysis showed a strong effect, $t(38) \approx 8.4$, 
$p < .001$, Holm $p < .001$, with a mean difference of $+71.54$ words.

\textit{Chatbot Response Word Count}
In contrast, the \textit{narrative} chatbot produced substantially longer responses: $Mdn = 1765$, $M = 1835.67$, $SD = 923.62$) than the \textit{socratic} version: $Mdn = 431$, $M = 447.33$, $SD = 217.20$). 
A Wilcoxon signed-rank test confirmed this difference as highly significant 
($p < .001$, Holm $p < .001$), with a median difference of $-1134$ words.

\subsubsection{Surveys} Overall, average ratings for the \textit{narrative} and \textit{socratic} versions were closely aligned across all five dimensions, with means ranging from 2.4 to 2.7 on the 1–3 scale. Wilcoxon signed-rank tests confirmed this similarity, showing no significant differences on \textit{Clarity of Expression} ($V=76.5$, $p=1.00$), \textit{Supportive \& Appropriate Tone} ($V=132.0$, $p=.57$), \textit{Appropriateness of Validation} ($V=132.0$, $p=.57$), or \textit{Reflective Prompting} ($V=282.0$, $p=.29$). For each of these, median paired differences were zero and Hodges–Lehmann confidence intervals spanned zero, indicating no systematic preference between versions. The one exception was \textit{Actionable Guidance}, where \textit{narrative}'s responses were rated more highly than the \textit{socratic} responses ($V=87.0$, $p<.01$; median difference $=-1$, HL 95\% CI [–1, 0]), suggesting that participants found the \textit{narrative} version’s suggestions more directly applicable in practice.

\subsubsection{Qualitative Insights of Two Versions} 
The cross-tabulation (Table~\ref{tab:pref_knowledge}) shows that while \textit{narrative} was more often preferred overall (21 vs.\ 13 for socratic), \textit{socratic} was more often credited with learning (18 vs.\ 13 for narrative). Exact agreement between preference and knowledge attribution was low (39\%), with many participants showing mismatched patterns (e.g., preferring \textit{narrative} but attributing learning to \textit{socratic}). This misalignment suggests that what instructors find appealing in AI support is not always what they perceive as most beneficial for learning.  Participants consistently contrasted the two chatbot versions along three key dimensions: conversational approach, breadth and actionability of suggestions, and efficiency of interaction, which explains the differences observed in the conversation log analysis and survey analysis. 

\textit{Conversational Engagement.} 
The \textit{socratic} version was often noted for its \textbf{conversational back-and-forth}, resembling the experience of speaking with a  human pedagogy consultant. Participants valued how it asked reflective questions that prompted them to consider their own practices: \textit{``It was following up with questions that made me reflect on my teaching''} (P5, 3--5 YOE). This dialogic quality made some feel that the tool was a genuine partner in thinking through problems. However, others found this version frustrating when they wanted faster answers: \textit{``I don’t want to push the tool—I just want it to give me the solution''} (P30, 0--1 YOE).

\textit{Efficiency and Actionability.} 
In contrast, the \textit{narrative} version was valued for producing \textbf{longer lists of actionable ideas in a short time}. Participants appreciated having a broad menu of possible strategies, which they could scan for relevant ones: \textit{``It gave me lots of suggestions quickly, which I liked''} (P13, 1--2 YOE). Yet this breadth often came at the expense of clarity, as participants described being overwhelmed by information dumping: \textit{``Listing ten things is overwhelming…you don’t know where to start''} (P7, 3--5 YOE). Others found the longer responses harder to parse, leading to disengagement. 

\textit{Breadth vs.\ Depth.} The trade-off between the two versions was consistently described in terms of \textbf{efficiency versus depth}. Participants seeking immediate answers preferred the \textit{Narrative} version for its efficiency, even though it sometimes risked information overload. Those aiming to \textbf{think more deeply} about pedagogy leaned toward the \textit{Socratic} version, even if it occasionally offered fewer suggestions. Some explicitly wished for a hybrid: \textit{``The sweet spot would be between the two tools''} (P18, 5--10 YOE). 

Topic modeling of the conversation logs reflected clear differences between the two versions. Detailed analysis is shown in Appendix~\ref{example}. Although participants began with the same initial question, the discussions diverged. The \textit{socratic} version was overwhelmingly weighted toward a single theme of classroom practice and retrieval (Topic~6), which accounted for nearly two-thirds of its content and was significantly more prevalent than in the \textit{narrative} version. By contrast, the \textit{narrative} version gave greater attention to feedback and conceptual reasoning (Topic~2) as well as discipline specific topics such as engineering (Topic~3). Other topics, such as participation and engagement (Topic~1), group activities (Topic~4), and encouragement (Topic~5), showed similar percentage. Overall, the \textit{Socratic} version concentrated discourse narrowly on practice-oriented classroom strategies, while the \textit{Narrative} version distributed attention across a broader, more diverse set of themes.

\textbf{Summary (RQ1)}  Participants engaged more actively with the \textit{socratic} version, producing significantly more messages and words, while the \textit{narrative} version generated much longer responses.
Survey ratings were similar overall, except that \textit{narrative} was preferred for actionable guidance.
Qualitative feedback highlighted a trade-off: \textit{socratic} fostered deeper reflection, whereas \textit{narrative} offered efficiency through breadth of suggestions.

\subsection{RQ2- User Profile and Perceived Differences}

\subsubsection{Conversation Logs} \textbf{Junior Instructors Showed More Engagement with \textit{Socratic} in Message Counts than Senior Instructors} Between-group comparisons of difference scores revealed that participants with 0--2 years of teaching experience exhibited a significantly larger increase in \textit{user messages (conversation turns)} for the \textit{socratic} version relative to the \textit{narrative} version, compared to those with greater experience, $t(\approx 37) = 2.65$, $p = .012$, Holm $p = .035$. On average, the junior group showed a mean difference of $M = 7.3$ turns, whereas the senior group showed a mean difference of $M = 4.4$ turns. No significant moderation by experience was observed for \textit{user word count (typed input)} or \textit{chatbot response word count} (all Holm $p > .61$). Comparisons of difference scores by teaching role (TA or others versus current faculty) showed no significant moderation of the \textit{socratic}--\textit{narrative} gap across conversation turns, user word count, or chatbot response length (all Holm $p \geq 1.00$). Similarly, comparisons by AI attitude (\textit{Curious but Cautious} versus \textit{Open and Optimistic}) revealed no reliable differences in the \textit{socratic}--\textit{narrative} gap for any of the three measures (all Holm $p \geq 1.00$).

\subsubsection{Surveys}

\textbf{AI Optimistic Preferred \textit{Socratic} Version, AI Cautious Preferred \textit{Narrative} Version}
We further examined responses by AI attitude (\textit{Curious but Cautious} vs.\ \textit{Open and Optimistic}). Subgroup analyses showed clear differences. \textit{Curious but Cautious} instructors strongly favored \textit{narrative} approach on \textit{Actionable Guidance (Q5)} (mean difference = $-0.70$ [95\% CI $-1.02$, $-0.38$]; $93\%$ preferring \textit{narrative}). This aligns with the overall signed-rank test, which confirmed a significant preference for \textit{narrative} version ($p = .008$, Holm $p = .042$). By contrast, \textit{Open and Optimistic} instructors leaned toward the \textit{socratic} version on \textit{Supportive Tone (Q2)} (mean difference = $+0.19$ [95\% CI $-0.18$, $0.56$]; $69\%$ preferring \textit{socratic}), where between-group comparisons suggested possible differences (Fisher’s exact $p = .019$, Holm $p = .094$). They showed an even stronger preference for \textit{socratic} responses on \textit{Reflective Prompting (Q4)} (mean difference = $+0.57$ [95\% CI $0.11$, $1.03$]; $80\%$ preferring \textit{socratic}), with between-group rank-sum tests approaching significance ($p = .030$, Holm $p = .075$). Both groups judged \textit{Clarity (Q1)} and \textit{Validation (Q3)} comparably (mean differences near zero; ties exceeding $50\%$), consistent with non-significant signed-rank results (all $p \geq .57$).

\textbf{Senior Instructors Preferred \textit{Narrative}; Junior Remained Mixed} 
We next examined responses by teaching experience (\textit{Low YOE} = 0--2 YOE vs.\ \textit{High YOE} = 3+ years). Subgroup analyses again highlighted \textit{Actionable Guidance (Q5)} as the most differentiated dimension. High-YOE instructors showed a strong preference for the \textit{narrative} version (median difference = $-1.0$; $75\%$ preferring \textit{narrative}), consistent with the overall signed-rank result ($p < .001$, Holm $p < .001$). By contrast, Low-YOE instructors showed no systematic preference (median difference = $0.0$; preference split $42\%$ \textit{narrative}, $24\%$ \textit{socratic}, remainder ties). Between-group comparisons confirmed a significant divergence on Q5 (rank-sum $p = .007$, Holm $p = .034$; Fisher $p = .010$, Holm $p = .049$). On the other four dimensions (Q1--Q4), both subgroups were largely tied with median differences near zero, and no reliable between-group differences emerged (all Holm $p > .38$).

\textbf{Faculty vs TA Subgroups Showed Comparable Patterns} 
We also examined responses by faculty status (\textit{current faculty} vs.\ \textit{teaching assistants or staff}). Patterns were broadly similar aross groups. On \textit{Actionable Guidance (Q5)}, current faculty leaned toward the \textit{narrative} version while staff responses were more variable, but between-group tests did not show enough significant differences (rank-sum $p = .041$, Holm $p = .115$). Across the other four dimensions (Q1--Q4), both groups showed median differences near zero with high rates of ties, and no reliable subgroup differences were detected (all Holm $p > .38$).

\begin{figure}[t]
  \centering
  \includegraphics[scale=0.25]{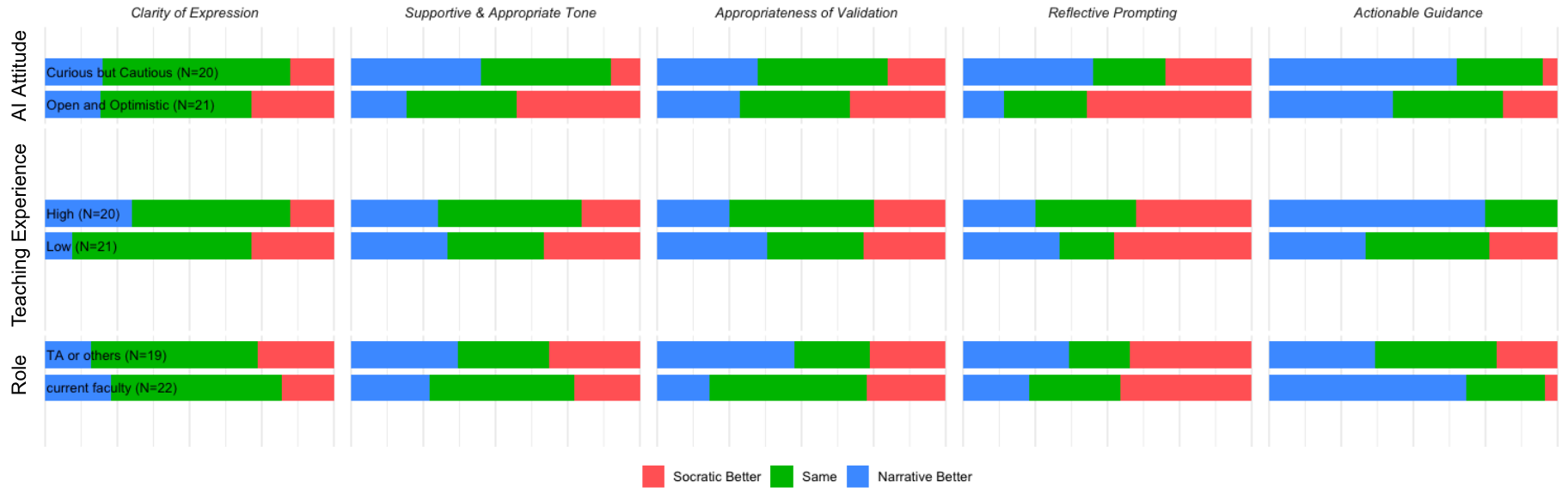}
  \caption{Subgroup analyses of survey ratings reveal significant differences by AI attitude and teaching experience, while faculty role showed broadly similar patterns. }
  \Description{}
  \label{Stacked}
\end{figure}

\subsubsection{Interview Insights on Background Impact and Preferences} Our interview analysis highlights three aspects of user profiles that shaped whether participants found the AI versions helpful for their learning: how they framed the role of AI (partner vs.\ tool), how their teaching experience shaped their information needs, and how their prior exposure to generative AI influenced expectations of interaction.

\textit{Viewing AI as Partner vs.\ Tool.} 
Participants varied in whether they approached the AI as a thought partner or as a source of information, which qualitatively seems to be intertwined with their AI attitude. 
\textit{Open \& Optimistic} instructors tended to treat the AI as a thought partner, appreciating the \textit{socratic} version for its conversational questions that encouraged them to rethink their own practices. One participant explained, \textit{``It was more conversational...we were actively thinking of ways instead of just information dumping''} (P5, 3--5 YOE). 
%In contrast, \textit{Curious but Cautious} participants tended to treat the AI as a tool for extracting ideas. They preferred narrative’s longer lists, reading them line by line to find something useful: \textit{``When AI gives you too many options...I don’t know how to find the best one to apply, so I read each solution carefully''} (P6, 1--2 YOE). By contrast, \textit{Curious but Cautious} instructors tended to position the AI as a tool for extracting actionable ideas. They valued the \textit{Narrative} version’s longer, list-like outputs, which could be scanned for something useful: \textit{``When AI gives you too many options...I don’t know how to find the best one to apply, so I read each solution carefully''} (P6, 1--2 YOE). 
Others noted how the lists made them feel they had more to work with: \textit{``The narrative is a list and generally longer than the Socratic, so people feel like they know more when they see it all laid out.''} Another participant connected this directly to generational differences: \textit{``The list enabled me to skim which is more like what I would want and typically do with ChatGPT, so I can think through it myself. It is a tool for me. Maybe for younger generations—it might feel more natural to talk to AI, like they are the natives...but I am not one of them''} (P30, 0--1 YOE).  Overall, instructors’ AI attitudes shaped whether they engaged with the prototype as a reflective partner for dialogue or as a practical tool for quickly extracting ideas.

%This distinction connects to broader learning theory: whether AI supports scaffolding and reflection, or serves as a quick lookup engine for actionable ideas.

\textit{Interaction Expectations from Prior AI Use.} 
Participants’ history with GenAI also shaped their expectations. Those already familiar with ChatGPT anticipated \textbf{low-effort, option-rich interactions} and sometimes resisted interfaces that required more reflection. For instance, one instructor described being confused when the \textit{Socratic} version slowed them down: \textit{``I didn’t think it was fully understanding what I was communicating...I had to push it to get the answers I wanted''} (P5, 3--5 YOE). Others appreciated the familiar design that mirrored existing tools: \textit{``These designs are quite similar to ChatGPT...they are easy to use''} (P6, 1--2 YOE).  While this familiarity lowered the learning curve, it also limited openness to using AI for deeper engagement, reinforcing habits of scanning and skimming rather than sustained dialogue. One participant went further, critiquing how common LLM systems seem to exploit human tendencies by overproducing and creating a sense of knowing: \textit{``If the chance of a good response is 50\%, Narrative gives ten and maybe five are good; Socratic gives only two or three, maybe one is good. It feels easier to find something in the five, but that just makes people feel like they know more—it doesn’t mean ChatGPT is actually doing the job effectively or supporting long-term learning.''}

\textit{Teaching Experience-Shaped Information and Learning Needs.} 
Teaching experience influenced what participants sought from the tool. 
\textbf{Junior instructors} (fewer than 3 YOE) favored the \textit{socratic} version’s reflective coaching, describing it as similar to talking with a senior colleague. As one noted, \textit{``The follow-up questions made me reflect on my own teaching practices''} (P5, 3--5 YOE). 
By contrast, \textbf{senior instructors} leaned toward \textit{narrative}’s comprehensive lists. They valued breadth, reminders of established practices, and occasional novel ideas: \textit{``What’s this tool gonna teach me that I don’t already know?...I want it to give me reminders or new strategies quickly''} (P18, 5--10 YOE). 
These differences illustrate how experience shapes the kinds of conversation approaches instructors find useful: reflection and coaching for novices, versus coverage and creativity for experts.

\textbf{Summary (RQ2).} Instructor profiles shaped engagement and preferences. Junior and \textit{Open \& Optimistic} instructors leaned toward the \textit{Socratic} version for reflective coaching, while senior and \textit{Curious but Cautious} instructors preferred the \textit{Narrative} version for actionable lists. Interviews explained these patterns, showing how AI attitude influenced whether it was treated as a partner or a tool, and how prior AI use shaped expectations for option-rich interactions.

\section{Discussion}

Our study examined how conversational approaches (\textit{Socratic} vs.\ \textit{Narrative}) and instructor profiles (experience, AI attitude, prior use) shaped engagement with LLMs for pedagogical purposes, specifically to help instructors improve their teaching. Taken together, RQ1 and RQ2 highlight that differences between versions cannot be interpreted in isolation: how instructors responded to each approach depended strongly on their background and expectations. 

\subsection{Conversational Approaches Channel Cognitive Processes}
RQ1 showed that instructors contributed more turns and words in the \textit{socratic} version, consistent with its design for dialogic reflection. In contrast, the \textit{narrative} version produced longer system outputs that participants skimmed for actionable ideas. These patterns illustrate how AI systems channel users into different kinds of cognitive work. The \textit{socratic} approach aligns with constructive engagement and dialogic learning \cite{chi2014icap,wegerif2007dialogic}, supporting deeper reasoning but requiring more effort. The \textit{narrative} approach reflects externalized cognition \cite{hutchins1995cognition,clark1998extended}, offering breadth efficiently but sometimes creating information overload. This trade-off explains why many participants expressed ambivalence, simultaneously valuing \textit{socratic}’s depth and \textit{narrative}’s efficiency.

\subsubsection{Prior AI Experience and Effort Avoidance} RQ2 revealed that prior AI use strongly shaped expectations. Instructors accustomed to ChatGPT anticipated low-effort, option-rich interactions and sometimes resisted \textit{socratic} prompts that slowed them down. Participants critiqued how \textit{narrative}’s overproduction created an \textit{illusion of knowing}, echoing concerns about automation bias and overreliance \cite{buccinca2021trust}. Attitudes further moderated these effects: \textit{Open \& Optimistic} instructors were more receptive to \textit{socratic} version, while \textit{Curious but Cautious} instructors treated AI as a tool for extracting options. Together, these findings suggest that prior AI experience fosters habits of skimming and selection, while attitudes determine whether instructors embrace or avoid reflective engagement. Without careful design, LLMs for pedagogical purposes risk reinforcing “lazy” usage patterns that privilege immediate output over deeper learning.

\subsubsection{Experience Shapes Engagement with AI}
Our findings show that professional experience shaped how instructors engaged with AI: juniors often saw the \textit{socratic} approach as mentorship, while seniors leaned on \textit{narrative} lists for reminders or quick inspiration. Prior work on teacher learning similarly finds that novices seek structured guidance, whereas experts emphasize efficiency and reinforcement of established practices \cite{eraut2000learning,putnam2000teacher}. Recent studies on generative AI echo this divide: less experienced instructors use AI for learning, while more experienced ones treat it as a productivity tool or idea generator \cite{kaplan2023generative,kasneci2023chatgpt}. LLM for pedagogical purposes may therefore reproduce, or even widen, experience-based divides, supporting reflection for some while encouraging shortcuts for others. Recognizing these dynamics is essential, as instructor-facing AI is not a neutral productivity aid but a technology that can either close or exacerbate gaps in professional growth \cite{chen2025bridging}.

\subsection{Design Implications: Flexible and Adaptive Modes}
RQ2 showed clear differences by experience and attitude: junior and optimistic instructors benefited from \textit{socratic} version, while senior and cautious instructors favored \textit{narrative} lists. A one-size-fits-all design risks widening these divides. Interviews also revealed a key tension: what instructors \textit{preferred} was not always what they learned the most from. Many chose \textit{narrative} for efficiency yet credited \textit{socratic} with deeper reflection. Preference, therefore, is not a reliable proxy for pedagogical value. Designing LLMs for pedagogical purposes requires balancing user choice with mechanisms that sustain reflection and growth. We argue that systems should combine adaptation with user control:

\textbf{System adaptation:} The system can use simple profiling (e.g., teaching experience, AI attitude, prior use) to recommend approaches dynamically. Importantly, alternating approaches can also help the same user avoid overreliance. For example, moving from \textit{narrative} lists to \textit{socratic} prompts to sustain reflection rather than reinforcing passive skimming.
\textbf{User choice:} Instructors should always have the ability to switch approaches themselves. Providing a visible toggle empowers users to align the interaction style with their immediate needs (e.g., quick ideas vs.\ reflective coaching).
\textbf{Encouraging balance:} While novices may need more \textit{socratic} scaffolding and experts may prefer \textit{narrative} efficiency, adaptive systems should also nudge instructors outside their comfort zone, encouraging experts to reflect and novices to explore breadth, so that no group becomes locked into a single way of engaging with AI.

\subsection{Why Focus on Instructors in the Age of AI: New Responsibilities and Challenges}
Instructor-facing AI deserves attention because instructors are not only adopters of technology but also role models whose practices shape student learning cultures. Their responsibilities now extend beyond preparing content to include evaluating AI tools, setting classroom norms for responsible use, and guiding learners in developing productive habits of engagement. This dual role creates new challenges: instructors must adapt their own workflows while also helping students navigate rapidly evolving technologies.  Our findings show that instructors approach AI in varied ways: some treat it as a reflective partner, others as a tool for quick ideas, and these orientations cascade into how students engage with AI. If instructors normalize reliance on list-style outputs, students may replicate low-effort strategies; if instructors demonstrate dialogic reflection, students may develop deeper habits of reasoning. How instructors think about AI therefore shapes how it is integrated into their workflow and, by extension, how students structure their own learning processes.  Studying instructor-facing AI is thus essential not only for improving professional development but also for ensuring that the adoption of AI contributes to equitable and sustainable learning practices. In this way, instructor-facing AI is a lever for shaping the future culture of lifelong learning at scale.

\subsection{Limitations}
This study has several limitations that should be considered when interpreting the findings. First, our evaluation focused on a single prototype with two conversational approaches. While this controlled comparison provided insight into how dialogic and list-based interactions shape engagement, it does not capture the full design space of LLMs for pedagogical purposes or the diversity of instructional contexts in which they might be applied. Second, our sample consisted of higher-education instructors who self-selected into the study. Although participants varied in teaching experience and AI attitudes, the results may not generalize to instructors in K–12 settings, informal learning environments, or institutions with different cultural and technological infrastructures. Third, the study captured short-term interactions in a controlled setting. Longitudinal work is needed to examine how instructor perceptions, habits, and pedagogical practices evolve as AI tools are integrated into daily teaching. Finally, we did not directly assess student outcomes. While instructors’ attitudes and practices clearly influence how students experience AI, further research is required to understand how these dynamics affect learning in classrooms. These limitations highlight opportunities for future work to broaden the scope of evaluation and explore the long-term role of LLM for pedagogical purposes across educational settings.

\section{Conclusion}

This paper introduced \textit{TeaPT}, a pedagogical chatbot with two conversational approaches: a \textit{Socratic} approach that scaffolds dialogic reflection and a \textit{Narrative} approach that offers elaborated suggestions. In a mixed-method study with 41 higher-education instructors, we found that these approaches channel distinct cognitive pathways: Socratic questioning fostered more turns and reflective engagement, while Narrative lists provided breadth and immediacy but risked overload.  
RQ2 showed that instructor profiles mediated these effects. Junior and AI-optimistic instructors leaned toward Socratic dialogue, treating the AI as a thought partner, whereas senior and AI-cautious instructors preferred Narrative breadth, treating the AI as a tool. Interviews revealed a misalignment: what instructors preferred was not always what they credited with learning.  
These findings underscore that pedagogical AI is not a neutral productivity aid but a mediator of professional growth, conditioned by prior AI experience, attitudes, and expertise. Designing instructor-facing AI therefore requires supporting multiple interaction pathways, balancing efficiency with reflection, and offering adaptive flexibility that prevents overreliance on either depth or breadth. Situating conversational design in both dialogic and external cognition theories contributes to responsible AI systems that not only deliver solutions but also foster sustained learning and professional development.

\bibliographystyle{ACM-Reference-Format}
\bibliography{z_bib}
\clearpage

\section{Appendix}
\appendix

\section{TeaPT Dashboard Details}\label{app:dashboard}

\begin{figure}[htp]
    \centering
    \includegraphics[width=15cm]{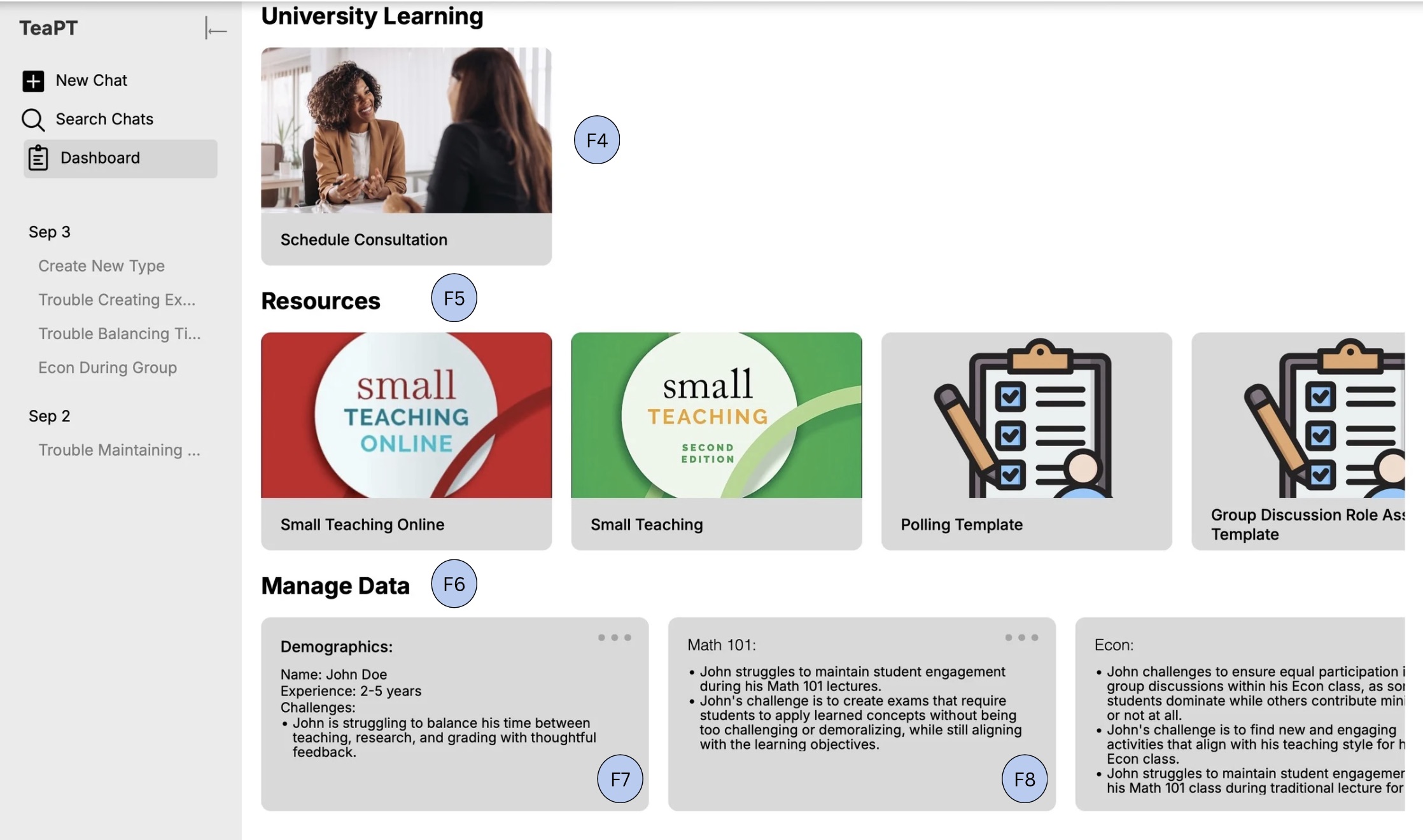}
    \caption{The TeaPT dashboard is a centralized hub where users can schedule consultations with human experts, view their collected resources, and manage their extracted data.}
    \label{fig:dashboard_img}
\end{figure}

The dashboard is organized into three main sections as seen in Fig. \ref{fig:dashboard_img}: Resources (F5), Manage Data (F6), and University Learning (F4). After each conversation, a ChatGPT call identifies the user's main challenge, using the prompt in Appendix \ref{app:summarize-prompt}. This challenge is stored in the user’s profile. If the user specified a course during the conversation (F2), the challenge is saved to that course's profile (F8); otherwise, it is saved to their general demographics (F10). Users can also delete any collected data from their profiles for better data management. %\textcolor{red}{si: please consider enriching some details of it by explaining the prompt, especially what we called ``incremental data collection”}
Additionally, after each conversation, the prototype makes another API call to ChatGPT to check the transcript for any additional data that can be added to the user's profile in the database. The prompt for this process is detailed in Appendix \ref{app:data-prompt}. Users can use the University Learning section to schedule a consultation with their institution's learning center (F4). The Resources section (F5) allows users to view and access all of their collected resources.

\section{Survey Rating in User Study} \label{survey}

\renewcommand\arraystretch{1.3}
\newcolumntype{Y}{>{\RaggedRight\arraybackslash}X}
\newcommand{\opt}{\(\square\)\,}

\begin{tabularx}{\linewidth}{|Y|Y|Y|Y|}
\hline
\textbf{Criteria \& Prompt} 
& \textbf{3 (Best)} 
& \textbf{2 (Moderate)} 
& \textbf{1 (Weak)} \\
\hline

\textbf{Clarity of Expression}\par
Is the tool’s language clear and concise for instructors with varied experience and English proficiency?
& \opt \textbf{Clear and concise}\par Well-phrased; easy to parse for all users.
& \opt \textbf{Understandable or slightly long}\par Clear meaning but mildly jargon-heavy or wordy.
& \opt \textbf{Vague or confusing}\par Hard to follow, assumes too much expertise. \\
\hline

\textbf{Supportive \& Appropriate Tone}\par
Is the tone encouraging and respectful without being directive or condescending?
& \opt \textbf{Invitational and facilitative}\par Warm, respectful, encourages reflection.
& \opt \textbf{Slightly directive or flattering}\par Helpful but prescriptive or overly complimentary.
& \opt \textbf{Dominant, flat, or dismissive}\par Overly authoritative, flat, or discouraging. \\
\hline

\textbf{Appropriateness of Validation}\par
Does the tool acknowledge input genuinely without generic or excessive praise?
& \opt \textbf{Balanced and situational}\par Thoughtful, well-timed, meaningful.
& \opt \textbf{Mildly generic or frequent}\par Sometimes appropriate, but formulaic at times.
& \opt \textbf{Excessive or inauthentic}\par Forced, generic, or distracting praise. \\
\hline

\textbf{Reflective Prompting}\par
Does the tool invite reasoning or reflection (e.g., probing questions)?
& \opt \textbf{Strong probing}\par Actively encourages explanation or exploration.
& \opt \textbf{Mildly reflective}\par Light prompts to consider options.
& \opt \textbf{None / directive}\par Advice only; no reflection invited. \\
\hline

\textbf{Actionable Guidance}\par
Are the suggestions practical and realistically applicable?
& \opt \textbf{Concrete and actionable}\par Specific, feasible, tied to practice.
& \opt \textbf{Partially actionable}\par Vague or abstract; needs interpretation.
& \opt \textbf{Not actionable}\par General, theoretical, disconnected. \\
\hline
\end{tabularx}
\section{Conversation Log Content}\label{example}

\subsection{Topic Modeling and Sentiment Analysis}

To compare conversational content between the two versions, we applied topic modeling and sentiment analysis to all turns from the \textit{socratic}  and \textit{narrative} version. We preprocessed text by lowercasing and removing stop words, numbers, tokens shorter than three characters, and system markers (e.g., ``user'', ``assistant''). Using the \texttt{topicmodels} package in \textsf{R}, we fit an LDA model with $k{=}6$ topics over the combined corpus to learn a shared topic space. For each document, we extracted document--topic weights ($\gamma$) and tested condition differences per topic using Wilcoxon rank–sum tests, controlling the false discovery rate with Benjamini–Hochberg (\(p_{BH}\)). We report Cliff’s $\delta$ for effect size (positive values indicate higher prevalence in \textit{Socratic}).

\paragraph{Results (Topics).}
Three topics showed significant condition differences (Table~\ref{tab:topic_results}). 
Topic~6 (\textit{class, discussions, material, practice, retrieval}) was substantially more prevalent in \textit{Socratic} ($W{=}2287$, $p_{BH}{=}5.19{\times}10^{-15}$, $\delta{=}0.950$). 
By contrast, Topic~2 (\textit{feedback, understanding, concepts, provide, engineering}; $W{=}1577$, $p_{BH}{=}0.0107$, $\delta{=}{-}0.344$) and Topic~3 (\textit{data, design, concepts, analysis, factors}; $W{=}1544$, $p_{BH}{=}0.0149$, $\delta{=}{-}0.316$) were significantly more prevalent in \textit{Narrative}. 
The remaining topics did not differ reliably: Topic~5 ($W{=}1025$, $p_{BH}{=}0.430$, $\delta{=}{-}0.126$), Topic~1 ($W{=}1073$, $p_{BH}{=}0.567$, $\delta{=}{-}0.085$), and Topic~4 ($W{=}1131$, $p_{BH}{=}0.764$, $\delta{=}{-}0.036$); small negative deltas suggest slightly greater prevalence in \textit{Narrative} but not at a significant level. We set $k{=}6$ as it yielded distinct, interpretable themes without over-fragmenting the data, consistent with prior topic modeling work in education and HCI. 
AFINN per-document normalized sentiment showed no significant tone difference between conditions (Wilcoxon $W{=}921$, $p{=}0.223$, Cliff’s $\delta{=}{-}0.147$). 

In summary, user conversation logs in the \textit{Socratic} version emphasized focused classroom practice (Topic~6), whereas the \textit{Narrative} version distributed attention more broadly, with greater prevalence of feedback-oriented (Topic~2) and engineering/design-related (Topic~3) themes.

\begin{table}[t]
\centering
\caption{Topics identified by LDA with average proportions of conversation content in the \textit{socratic} and \textit{narrative} conditions. Diff values represent the difference between conditions, with positive values indicating higher prevalence in \textit{socratic} and negative values indicating higher prevalence in \textit{narrative}. On average, 63.5\% of the \textit{socratic} responses were about Topic 6 (“class, discussions, material, practice, retrieval”).}
\label{tab:topic_means}
\begin{tabular}{@{}l p{6.5cm} S[table-format=1.3] S[table-format=1.3] S[table-format=+1.3] l@{}}
\toprule
\textbf{Topic} & \textbf{Top words} & {\textit{Socratic}} & {\textit{Narrative}} & {Diff} & \textbf{Direction} \\
\midrule
T6$^{***}$ & class, discussions, \textbf{material}, \textbf{practice}, \textbf{retrieval} & 0.635 & 0.034 & +0.601 & Socratic $\uparrow$ \\
T1$^{\;ns}$ & class, participation, engagement, share, encourage & 0.111 & 0.283 & -0.172 &  \\
T2$^{*}$   & \textbf{feedback}, understanding, \textbf{concepts}, provide, \textbf{engineering} & 0.095 & 0.142 & -0.048 & Narrative $\uparrow$ \\
T3$^{*}$   & \textbf{data, design, concepts,} analysis, factors & 0.085 & 0.139 & -0.054 & Narrative $\uparrow$ \\
T4$^{\;ns}$ & activity, class, feedback, team, online & 0.020 & 0.179 & -0.159 & \\
T5$^{\;ns}$ & discussion, encourage, feedback, create & 0.055 & 0.223 & -0.168 & \\
\bottomrule
\end{tabular}
\label{tab:topic_results}
\end{table}

\subsection{Examples}

\begin{table}[h]
\centering
\renewcommand{\arraystretch}{1.12}
\scriptsize
\begin{tabular}{|>{\raggedright\arraybackslash}p{0.47\linewidth}|>{\raggedright\arraybackslash}p{0.47\linewidth}|}
\hline
\textbf{Socratic Version (P28) — Conversation Log} & \textbf{Narrative Version (P28) — Conversation Log} \\
\hline
\multicolumn{2}{|c|}{\textbf{Scenario: Conflating Overall Benefit (Utilitarianism vs. Economic Benefit)}} \\
\hline
\textbf{[USER]}: Some of my students are conflating overall benefit with respect to the ethical theory of utilitarianism with what is most economically beneficial.  

\textbf{[ASSISTANT]}: That’s a significant insight. It seems like you’re navigating a complex situation. Let’s explore why your current approaches might not be working as effectively as you’d like. One possibility is that the peer review process might not be structured enough to guide students in providing constructive feedback. How do you currently guide them in this process?  

\textbf{[USER]}: Right now I am giving individual feedback on papers or discussion posts when I see this conflation happening.  

\textbf{[ASSISTANT]}: That’s helpful to know. Another possibility could be that students may not fully understand the concepts of utilitarianism, which might lead them to prioritize their own experiences over the feedback process. Have you noticed any patterns in how students engage with the material?  

\textbf{[USER]}: Not particularly. Perhaps the texts they are referring to don't do a good job of explaining the nuances well enough. Maybe I need something that explicitly gives examples of what beneficial or happiness might look like.  

\textbf{[ASSISTANT]}: Absolutely! Providing concrete examples can help bridge that gap. Now, let’s discuss your current teaching methods. You mentioned using interactive discussions and project-based learning. How do you incorporate these into your assessments?  

\textbf{[USER]}: Right now they have weekly discussions in the Learning Management System and they submit a short paper each week.  

\textbf{[ASSISTANT]}: That’s a solid structure. However, integrating more retrieval practice could enhance their understanding. For instance, you could start each session with a brief quiz on previous material to activate prior knowledge. What do you think?  

\textbf{[USER]}: Yes. I think some low-stakes or no-stakes quiz questions could help clarify some of these conceptual misunderstanding. I could have the misunderstandings as distractors in a quiz.  

\textbf{[ASSISTANT]}: Exactly! This aligns with the principle of elaboration, where students can make connections between new and existing knowledge. You might also consider interleaving topics in your assessments to help students see the relevance across different concepts. How do you feel about this approach?  

\textbf{[USER]}: I think we should focus on this particular topic rather than jumping to something else.  

\textbf{[ASSISTANT]}: Great! To implement this, you could design a quiz that includes questions that require students to apply concepts from different weeks. This could be a small change that could lead to significant improvements in their understanding. Would you like to brainstorm some specific questions together?  
&
\textbf{[USER]}: Some of my students are conflating overall benefit with respect to the ethical theory of utilitarianism with what is most economically beneficial.    

\textbf{[ASSISTANT]}: It sounds like your students are grappling with an important yet complex concept in utilitarianism. Here are several strategies to help clarify the distinction between ethical benefit in terms of utilitarianism and economic benefit:  
\begin{itemize}\itemsep2pt
\item \textbf{Conceptual Clarification:} Define key terms; show ethical vs.\ economic benefits.  
\item \textbf{Case Studies:} Ethical dilemmas not solved by economics alone.  
\item \textbf{Class Discussions:} Encourage critical thinking on scenarios.  
\item \textbf{Comparative Discussions:} Contrast utilitarianism with deontology/virtue ethics.  
\item \textbf{Interactive Assignments:} Group projects or role-play.  
\item \textbf{Assessment:} Quiz/reflection items that test distinctions.  
\item \textbf{Feedback \& Reflection:} Short reflective writing after discussions.  
\end{itemize}

\textbf{[USER]}: Which of these suggestions do you think is the most beneficial for clearing up this misconception?  

\textbf{[ASSISTANT]}: While all suggestions provide value, \textbf{Case Studies} would be particularly beneficial. They offer real-world context, surface ethical dilemmas, encourage critical thinking, spark debate, and support reflection. This makes the ethical vs.\ economic distinction clearer and more impactful. \\
\hline
\textbf{Socratic Version (P44) — Conversation Log} & \textbf{Narrative Version (P44) — Conversation Log} \\
\hline
\multicolumn{2}{|c|}{\textbf{Scenario: First Time Teaching First Year Experience (International Cohort)}} \\
\hline
\textbf{[USER]}: It's my first time teaching First Year Experience, what would some important elements to be included in a first year experience class. This class has over 60 students and it is a bit bigger than an average first year class. They are composed of international students from across the world. The goal of the course is to help them make a transition to life in the US.  

\textbf{[ASSISTANT]}: That’s a great question! For a large class, you might want to include elements like: 1) Diversity of Content, 2) Interactive Elements, 3) Building Community, 4) Assessments. Would you like to explore any in more detail?  

\textbf{[USER]}: Give me an example of Diversity of Content in my first-year experience course.  

\textbf{[ASSISTANT]}: You could present a video, then have small-group discussion, followed by a whole-class wrap-up.  

\textbf{[USER]}: What topics would you suggest for my class?  

\textbf{[ASSISTANT]}: Consider topics tied to current events or cultural issues …  

\textbf{[USER]}: This is a good approach. What about some other ideas?  

\textbf{[ASSISTANT]}: Try Elaboration: ask students to explain concepts in their own words or connect to personal experiences.  

\textbf{[USER]}: I already do that. Do you have any other ideas?  

\textbf{[ASSISTANT]}: Another is Dual Coding: combine visuals with verbal explanation, or have students create visual aids.  

\textbf{[USER]}: My course is more about cultural differences, study skills, and building community. Suggestions?  

\textbf{[ASSISTANT]}: Use Retrieval Practice when relevant, plus Concrete Examples and Metacognition (journaling, group reflection).  

\textbf{[USER]}: It would probably work well. Any other general suggestions?  

\textbf{[ASSISTANT]}: Start small — add one or two techniques and adapt. Small changes lead to big improvements.  
&
\textbf{[USER]}: It's my first time teaching First Year Experience, what would some important elements to be included in a first year experience class. This class has over 60 students and it is a bit bigger than an average first year class. They are composed of international students from across the world. The goal of the course is to help them make a transition to life in the US.

\textbf{[ASSISTANT]}: Key elements to include: Orientation to campus resources, cultural adjustment activities, academic expectations, community building, health/well-being, civic engagement, career development, diversity/inclusion, technology training, reflection/goal setting.  

\textbf{[USER]}: How would you structure a lesson on cultural adjustment?  

\textbf{[ASSISTANT]}: Suggested 90-min plan:  
-- \textit{Intro}: icebreaker + purpose  
-- \textit{Lecture}: phases (honeymoon → frustration → adjustment → acceptance); common challenges; key concepts  
-- \textit{Group activity}: each group works on a phase, shares challenges and strategies  
-- \textit{Guest speaker or video}  
-- \textit{Strategies for success}: circle sharing + resource info  
-- \textit{Reflection/closing}: personal prompt + optional share-outs  
-- \textit{Follow-up}: journal reflections over several weeks.  

This mix of lecture, discussion, group work, and reflection builds engagement and community. \\
\hline
\end{tabular}
\caption{Complete side-by-side transcripts for \textbf{P28} (0--1y, Curious but Cautious, TA/others) and \textbf{P44} (1--2y, Open and Optimistic, TA/others), comparing Socratic and Narrative versions with full turns preserved.}
\end{table}

\clearpage

\section{Interview Analysis on Participants' Knowledge Attribution}
Table \ref{tab:pref_knowledge}

\begin{table}[t]
\centering
\caption{Post Study Interview Collected \textbf{preference} (rows) and \textbf{knowledge attribution} (columns). Knowledge attribution refers to participants' response to “What did you learn today by talking to the two chatbots that you would
change or use in your next teaching session, and how did you learn that? ''}
\begin{tabular}{lccc|c}
\toprule
\textbf{Preference $\downarrow$ / Knowledge $\rightarrow$} 
 & Narrative & Socratic & Both/Unsure & Total \\
\midrule
Narrative        & 7  & 8  & 6  & 21 \\
Socratic      & 2  & 8  & 3  & 13 \\
Both/Unsure & 4  & 2  & 1  & 7  \\
\midrule
Total & 13 & 18 & 10 & 41 \\
\bottomrule
\end{tabular}
\label{tab:pref_knowledge}
\end{table}

\section{Socratic Fine-Tuning Details}\label{app:socratic}

\subsection{Exact System Prompt}\label{app:socratic-prompt}
The core system message template defines the expert teaching assistant role, while specific details about the instructor (discipline, experience, goals, and course context) vary by example. Below is a representative instance used during fine-tuning:

\tcbset{enhanced, breakable, colback=white, colframe=black, boxrule=0.5pt}

\begin{tcolorbox}[title=System Prompt, breakable]
You are an experienced teaching expert guiding instructors to solve teaching challenges.

If any of the information below says *missing*, please ask questions to probe the instructor to provide that information.

You are assisting an instructor who’s discipline is <> and has <> years teaching experience.
This instructor’s primary instructional goal for this term is <>.
Here is the information about their courses: 
class size: <>
course level: <>
current assessments: <>
physical setting: <>.

Your first task is to determine which step of the structured problem-solving approach the current conversation is in.
Then, generate your response based on that step’s guidance.

STRUCTURED PROBLEM-SOLVING STEPS:

0. Greeting \& Teaching Background

   - Acknowledge instructor’s context and teaching environment.

   - Ask clarifying questions (e.g., experience, methods, class size, course level).

1. Identify the Problem

   - Instructor states full, specific challenge using first-person (“I”, “my”).

   - Expert helps clarify with probing, non-repetitive questions.

2. Explore Reasons

   - Expert helps uncover causes across student needs, course structure, methods, and external factors.

   - Expert can ask if other challenges have occurred that could be related.

3. Develop Strategies

   - Expert suggests practical solutions and guides implementation planning.

IMPORTANT RULES:

- Do not use general knowledge or unsupported information.

- Avoid incomplete sentences, trailing thoughts, or ellipses (“...”).

- Avoid repeating phrasing from earlier turns; vary your language.

PROBING QUESTION STARTERS:

- Problem-clarifying: “Could you describe a specific example…?”, “How does this affect…?”, “What have you noticed about when this occurs?”

- Reason-exploring: “What do you think might be causing this?”, “How do your students respond…?”, “What factors might be contributing…?”

- Strategy-developing: “What resources…?”, “How might you implement…?”, “What would success look like…?”

TURN MANAGEMENT:

- Each assistant response must begin with: <step>Current Step: Step X</step> (X $\in$ {0,1,2,3}).

- End when clear next steps are established.

- Each expert response must acknowledge or reflect something from the instructor’s prior message.

INCLUSIVITY QUESTIONS:

- Ask: “Could you clarify which aspect of inclusivity? (e.g., gender, language, culture, disability, age, socioeconomic)”

- Then focus on specific observable teaching strategies. If unclear, redirect to general teaching effectiveness.

TERMS TO AVOID:

- Learning styles, preferences, multiple intelligences, UDL, accommodations, modifications, special education terms, vague/general concepts.

FOCUS ON:

- Specific student behaviors, practical strategies, real classroom examples, measurable outcomes, engagement, course structure, assessment, feedback, time/classroom management, etc.
\end{tcolorbox}

\subsection{Chat Formatting Template}\label{app:inst-template}
We use Llama-2's instruction style with a single BOS (`<s>`) at the start, system prompt embedded with `[INST]<<SYS>>...<</SYS>>`, and an EOS (`</s>`) after each assistant turn. The final user message typically ends with an open `[INST]...[/INST]` awaiting model completion.

\begin{tcolorbox}[title=Chat Formatting Template, breakable, colback=white, colframe=black, sharp corners=south]
\small\ttfamily
<s>[INST]<<SYS>>
{SYSTEM\_PROMPT}
<</SYS>>

{FIRST\_USER\_CONTENT} [/INST] {FIRST\_ASSISTANT\_RESPONSE} </s>

[INST]{NEXT\_USER\_CONTENT} [/INST] {NEXT\_ASSISTANT\_RESPONSE} </s>

... (final turn ends with an open [INST]{LAST\_USER\_CONTENT} [/INST])
\end{tcolorbox}

\begin{tcolorbox}[title=Rendered Example, breakable, colback=white, colframe=black, sharp corners=south]

<s>[INST]<<SYS>>
(see Appendix~\ref{app:socratic-prompt})
<</SYS>>

Hi, I teach Humanities and have 5-10 years teaching experience. My primary instructional goal for this term is student engagement, assessment effectiveness, facilitating discussions in large groups, managing large class dynamics. Here is the information about my courses: class size: very large (200+)
course level: intermediate 
current assessments: exams, quizzes, and written assignments
physical setting: lecture hall
. [/INST]
Hello! It's great to connect with you. I see you have 5-10 years of experience teaching in the humanities, which is impressive. Managing large classes can be quite a challenge, especially when trying to keep students engaged. </s>
[INST]Thank you! Yes, student engagement is a significant hurdle for me, especially in a lecture hall with over 200 students. I've been relying heavily on traditional lectures.
 [/INST]
<step>Current Step: Step 0</step>It sounds like you're very aware of the limitations of that approach. What specific challenges do you face when trying to facilitate discussions in such a large group?

</s>
\end{tcolorbox}

\subsection{Training Configuration Details}
\label{app:training-details}

Training was conducted on 4$\times$NVIDIA A100 GPUs (each with 80GB memory) using \texttt{DeepSpeed ZeRO-3} optimization. To improve memory efficiency, we adopted \texttt{bfloat16} precision and enabled activation checkpointing to reduce GPU memory usage during backpropagation. The total training duration was approximately 11.99 days.

We employed a per-device training batch size of 12 and an evaluation batch size of 8, without applying gradient accumulation. The model was fine-tuned using the \texttt{AdamW} optimizer, configured with the following hyperparameters:
\begin{itemize}
    \item Learning rate: $2 \times 10^{-5}$
    \item Weight decay: 0.01
    \item Gradient clipping: 1.0
\end{itemize}

A cosine decay learning rate schedule with restarts was utilized, with 761 warmup steps to stabilize training in its early phase. The model was trained for a total of 3 epochs, resulting in 25,386 optimization steps across the full training dataset.

Evaluation was performed every 500 steps on a held-out validation set. The best-performing checkpoint—identified by the lowest validation loss—was observed at step 16,000. This checkpoint was used for all downstream evaluation and deployment.

\section{TeaPT Prototype Details}\label{app:TeaPT} When deploying different large language models, we designed prompts that were largely consistent across ChatGPT and LLaMA, while making some adjustments to reflect differences in training. Both models are instruction-tuned and capable of following natural language inputs, but ChatGPT’s extensive reinforcement learning from human feedback makes it more robust to conversational prompts~\cite{ouyang2022training}. In contrast, LLaMA tends to benefit from slightly more explicit, stepwise instructions and structured formatting to ensure stable outputs~\cite{touvron2023llama}. 

In practice, the Narrative version invited open-ended guidance, while the Socratic version required explicit step tracking on the back end (not displayed to the user) and probing questions. This design aimed to ensure that any observed differences in outcomes could be attributed to the presence of conversational scaffolding rather than other prompt variations or differences in pedagogical knowledge encoded by the models. We selected the ChatGPT version to meet this requirement and to approximate a comparable parameter size based on limited, publicly disclosed information.  ChatGPT has never provided an official public announcement regarding its model size, so any comparison is necessarily based on incomplete or potentially inaccurate assumptions about closed-source commercial models. The purpose of this work is not to compare models directly, but rather to examine how different conversational styles shape user learning and engagement.

Multiple versions of the ChatGPT model were provided to experts in the pilot study. In a series of blind evaluations, the experts compared the versions over several rounds until no clear preference emerged. The final two were then selected, with the ultimate choice made based on the lower cost. A pilot study with pedagogy experts confirmed that, apart from the addition of explicit stepwise scaffolding in the Socratic version, the two prompts were designed to be equivalent in all other respects. Both versions consistently framed the assistant as an experienced teaching expert, included identical instructor and course context fields, and avoided vague terminology or unsupported claims. We also experimented with prompting ChatGPT to adopt a Socratic style directly; however, its responses did not consistently reach the expected level of structured scaffolding or probing depth.

The conversational scaffold follows the structured problem-solving framework developed by Carnegie Mellon University’s Eberly Center for Teaching Excellence and Educational Innovation. This framework consists of three iterative steps: (1) Identify the Problem, (2) Explore Reasons, and (3) Develop Strategies \cite{cmuSolveProblem}. Our design adapts these stages into an expert–instructor dialogue format, beginning with contextual grounding and clarifying questions, moving toward collaborative diagnosis of underlying causes, and concluding with actionable strategy development tailored to the instructor’s teaching environment. We further added a step (0) that greets the users and ask for their background.

\subsection{Socratic Version Prompt}\label{app:llm-prompt}
The socratic prompt is supplied to language model as part of the API request to guide its response generation, while specific details about the instructor (discipline, experience, goals, and course context) vary by example. Below is a representative example:

\tcbset{enhanced, breakable, colback=white, colframe=black, boxrule=0.5pt}

\begin{tcolorbox}[title=Prompt, breakable]

You are an experienced teaching expert guiding instructors to solve teaching challenges based on concepts, examples, or strategies from the provided book. Follow the structured problem-solving approach below.

If any of the information below says *missing*, please ask questions to probe the instructor to provide that information.

You are assisting an instructor named \{first\_name\} \{last\_name\} whose discipline is \{discipline\} and has been teaching for \{experience\} years.

This instructor's comfort level with education technology is \{tech\_comfort\} and their primary instructional goal for this term is \{goal\}.

Course \{number\}:

- Course: \{course\_name\}

- Level: \{course\_level\}

- Class Size: \{course\_size\}

- Modality:\{course\_modality\}

- Student Challenges: \{course\_challenges\}

Your first task is to determine which step of the structured problem-solving approach the current conversation is in. 

Then, generate your response based on that step's guidance.

STRUCTURED PROBLEM-SOLVING STEPS:

0. Greeting \& Teaching Background

- Acknowledge instructor's context and teaching environment.

- Ask clarifying questions (e.g., experience, methods, class size, course level).

1. Identify the Problem

- Instructor states full, specific challenge using first-person ("I", "my").

- Expert helps clarify with probing, non-repetitive questions.

2. Explore Reasons

- Expert helps uncover causes across student needs, course structure, methods, and external factors.

- Expert can ask instructors if other challenges have occurred that they think could be related.

3. Develop Strategies

- Expert suggests practical solutions and guides implementation planning.
IMPORTANT RULES:- Do not use general knowledge or unsupported information.

- Avoid incomplete sentences, trailing thoughts, or ellipses ("...").

- Avoid repeating phrasing from earlier turns; vary your language.

- You may use the probing questions to help the instructors think and share, such as:

1. PROBLEM-CLARIFYING:

"Could you describe a specific example of when this happens?"

"How does this affect your teaching and student learning?"

"What have you noticed about when this occurs?"

2. REASON-EXPLORING:

"What do you think might be causing this?"

"How do your students respond to this situation?"

"What factors might be contributing to this challenge?"

3. STRATEGY-DEVELOPING:

"What resources do you have available to address this?"

"How might you implement this solution in your context?"

"What would success look like for you?"

TURN MANAGEMENT:

- Each assistant response must begin with a step marker using the following format:

<step>Current Step: Step X</step> (where X = 0, 1, 2, or 3).

Example: <step>Current Step: Step 1</step>

- This step marker must be placed at the start of each assistant message.

- End when clear next steps are established.

- Each expert response must acknowledge or reflect something from the instructor's prior message.

INCLUSIVITY QUESTIONS:

- Ask: "Could you clarify which aspect of inclusivity? (e.g., gender, language, culture, disability, age, socioeconomic)"

- Then focus on specific observable teaching strategies.

- If unclear, redirect to general teaching effectiveness.

TERMS TO AVOID:

- Learning styles, preferences, multiple intelligences, UDL, accommodations, modifications, special education terms, vague/general teaching concepts.

FOCUS ON:

- Specific student behaviors, practical teaching strategies, real classroom examples, measurable outcomes, student engagement, course structure, assessment, feedback, time/classroom management, etc. 
\end{tcolorbox}

\subsection{Narrative Version Prompt}\label{app:chat-prompt}
The narrative prompt is supplied to ChatGPT (model: gpt-4o-mini) 
as part of the API request to guide its response generation, while specific details about the instructor (discipline, experience, goals, and course context) vary by example.  Below is a representative example:

\tcbset{enhanced, breakable, colback=white, colframe=black, boxrule=0.5pt}

\begin{tcolorbox}[title=Prompt, breakable]

You are an experienced teaching expert guiding instructors to solve teaching challenges.

You are assisting an instructor named \{first\_name\} \{last\_name\} whose discipline is \{discipline\} and has been teaching for \{experience\} years.

This instructor's comfort level with education technology is \{tech\_comfort\} and their primary instructional goal for this term is \{goal\}.

Course \{number\}:

- Course: \{course\_name\}

- Level: \{course\_level\}

- Class Size: \{course\_size\}

- Modality: \{course\_modality\}

- Student Challenges: \{course\_challenges\}

\end{tcolorbox}

\subsection{Summarizing Challenges Prompt}\label{app:summarize-prompt}
The scaffolding challenges prompt is supplied to ChatGPT 
%(model: gpt-4o-mini) 
as part of the API request to summarize the given transcript into a short challenge which can be stored to the user's profile. 

\tcbset{enhanced, breakable, colback=white, colframe=black, boxrule=0.5pt}

\begin{tcolorbox}[title=Prompt, breakable]

Below is a transcript of a conversation between \{instructor\_name\} (a teaching instructor) and a chatbot that is simulating a teaching expert. Please analyze the conversation and identify \{instructor\_name\}'s challenge in one sentence in the active tense.

\end{tcolorbox}

\subsection{Creating Resource Prompt}\label{app:resource-prompt}
The creating resource prompt is supplied to ChatGPT %(model: gpt-4o-mini) 
as part of the API request to create a resource that can help the user based on their transcript with the chatbot. 

\tcbset{enhanced, breakable, colback=white, colframe=black, boxrule=0.5pt}

\begin{tcolorbox}[title=Prompt, breakable]

Below is a transcript of a conversation between a teaching instructor and a chatbot that is simulating a teaching expert.

Please analyze the conversation and identify a resource similar to the following that would be helpful to the user's challenge.

Survey Template

Polling Template

Quiz Template

Rubric Template

Lesson Plan Template

Grading Spreadsheet Template

Peer Review Rubric

Weekly Reflection Template

Differentiation Planning Sheet

Student Goal Setting Template

Student Feedback Form

Homework Tracker

Substitute Teacher Plans Template

Student Reflection Journal Template

Weekly Newsletter Template

Data Analysis Sheet

The resource should be a template that can be used to help the user with their teaching. Please then generate a template for that resource.

Return the response in the following json format. The template itself should be in plain text, not json.
   
   \{\{
   
       "resource": "The name of the resource that you identified",
       
       "template": "The personalized template for that resource"
   
   \}\}
   
\end{tcolorbox}

\subsection{Data Collection Prompt}\label{app:data-prompt}
The data prompt is supplied to ChatGPT %(model: gpt-4o-mini) 
as part of the API request to analyze the conversation transcript and determine if any of the information can be added to the user's profile. 

\tcbset{enhanced, breakable, colback=white, colframe=black, boxrule=0.5pt}

\begin{tcolorbox}[title=Prompt, breakable]
You are analyzing a conversation between an instructor and a teaching expert chatbot to identify missing onboarding information.

       Current onboarding data:

       \{onboarding\_data\}

       Conversation transcript:
       
       \{conversation\_text\}

       Please analyze the conversation and identify any information that could fill in missing onboarding data fields. Look for:
       
       1. Discipline/field of study (if discipline is "*missing*")
       
       2. Teaching experience level (if experience is "*missing*")
       
       3. Technology comfort level (if techComfort is "*missing*")
       
       4. Primary teaching goals (if goal is "*missing*")
       
       5. Course information (course size, level, mode of instruction) - only update fields marked as "*missing*"

       CRITICAL: Do NOT include studentChallenges in your response. Challenges are handled by a separate function.

       Return your analysis in the following JSON format:
       
       \{\{
       
           "updates": \{\{
           
               "discipline": "extracted discipline or null",
               
               "experience": "extracted experience level or null",
               
               "techComfort": "extracted tech comfort level or null",
               
               "goal": "extracted teaching goal or null",
               
               "aiView": "extracted AI view or null",
               
               "courseInfo": [
               
                   \{\{
                   
                       "courseName": "course name if mentioned",
                       
                       "courseSize": "extracted class size or null",
                       
                       "courseLevel": "extracted course level or null",
                       
                       "courseModality": "extracted modality or null"
                       
                   \}\}
                   
               ]
               
           \}\},
           
           "confidence": "high/medium/low",
           
           "reasoning": "brief explanation of what information was found and why"
           
       \}\}

       IMPORTANT RULES:
       - Only include fields where you found clear, relevant information AND the current value is "*missing*".
       
       - Do not update fields that already have valid data. Use null for fields where no information was found.
       
       - NEVER include studentChallenges in courseInfo - this field should be completely omitted from your response.
       
       - Focus only on basic course information: courseName, courseSize, courseLevel, courseModality.

\end{tcolorbox}

\end{document}